\documentclass[twocolumn,trackchanges]{aastex631}

\usepackage{multirow}
\usepackage{CJK}
\usepackage{enumerate}
\usepackage{longtable}
\usepackage{threeparttablex}
\usepackage{tablefootnote}

\usepackage{appendix}
\usepackage{float}
\usepackage{amsmath}

\newcommand{\comm}{\textcolor{black}}
\newcommand{\ping}{\textcolor{black}}
\newcommand{\rep}{\textcolor{black}}

\begin{document}
\begin{CJK*}{UTF8}{gbsn}
\title{GTC optical/NIR upper limits and NICER X-ray analysis of SGR~J1935+2154 for the outburst in 2022}

\correspondingauthor{Ping Zhou, Bin-Bin Zhang, Alberto~J.~Castro-Tirado}
\email{pingzhou@nju.edu.cn; bbzhang@nju.edu.cn; ajct@iaa.es}

\author[0000-0001-5684-0103]{Yi-Xuan Shao}
\affil{School of Astronomy and Space Science, Nanjing University, 163 Xianlin Avenue, Nanjing 210023, People's Republic of China}
\affil{Key Laboratory of Modern Astronomy and Astrophysics, Nanjing University, Ministry of Education, Nanjing 210023, People's Republic of China}

\author[0000-0002-5683-822X]{Ping Zhou}
\affil{School of Astronomy and Space Science, Nanjing University, 163 Xianlin Avenue, Nanjing 210023, People's Republic of China}
\affil{Key Laboratory of Modern Astronomy and Astrophysics, Nanjing University, Ministry of Education, Nanjing 210023, People's Republic of China}

\author[0000-0002-0584-8145]{Xiang-Dong Li}
\affil{School of Astronomy and Space Science, Nanjing University, 163 Xianlin Avenue, Nanjing 210023, People's Republic of China}
\affil{Key Laboratory of Modern Astronomy and Astrophysics, Nanjing University, Ministry of Education, Nanjing 210023, People's Republic of China}

\author[0000-0003-4111-5958]{Bin-Bin Zhang}
\affil{School of Astronomy and Space Science, Nanjing University, 163 Xianlin Avenue, Nanjing 210023, People's Republic of China}
\affil{Key Laboratory of Modern Astronomy and Astrophysics, Nanjing University, Ministry of Education, Nanjing 210023, People's Republic of China}

\author[0000-0003-2999-3563]{Alberto Javier Castro-Tirado}
\affiliation{Instituto de Astrof\'isica de Andaluc\'ia (IAA-CSIC), Glorieta de la Astronom\'ia s\/n, E-18008, Granada, Spain}
\affiliation{Departamento de Ingenier\'ia de Sistemas y Autom\'atica, Escuela de Ingenier\'ias, Universidad de M\'alaga, C\/. Dr. Ortiz Ramos s\/n, E-29071, M\'alaga, Spain}

\author[0000-0002-3386-7159]{Pei Wang}
\affil{National Astronomical Observatories, Chinese Academy of Sciences, Beijing 100101, China}

\author[0000-0003-3010-7661]{Di Li}
\affil{Department of Astronomy, Tsinghua University, Beijing 100084, People's Republic of China}
\affil{National Astronomical Observatories, Chinese Academy of Sciences, Beijing 100101, China}

\author[0000-0003-3047-607X]{Zeng-Hua Zhang}
\affil{School of Astronomy and Space Science, Nanjing University, 163 Xianlin Avenue, Nanjing 210023, People's Republic of China}
\affil{Key Laboratory of Modern Astronomy and Astrophysics, Nanjing University, Ministry of Education, Nanjing 210023, People's Republic of China}

\author[0000-0002-2420-5022]{Zi-Jian Zhang}
\affiliation{Department of Astronomy, School of Physics, Peking University, Beijing 100871, People's Republic of China}
\affiliation{Kavli Institute for Astronomy and Astrophysics, Peking University, Beijing 100871, People's Republic of China}
\affil{School of Astronomy and Space Science, Nanjing University, 163 Xianlin Avenue, Nanjing 210023, People's Republic of China}

\author[0000-0002-7400-4608]{You-Dong Hu}
\affiliation{Osservatorio Astronomico di Brera (INAF-OAB), Via Brera, 28, 20121 Milano MI, Italy}
\affiliation{Instituto de Astrof\'isica de Andaluc\'ia (IAA-CSIC), Glorieta de la Astronom\'ia s/n, E-18008, Granada, Spain}

\author{Shashi B. Pandey}
\affiliation{Aryabhatta Research Institute of Observational Sciences (ARIES), Manora Peak, Nainital-263002, India}

\begin{abstract}

The Galactic magnetar SGR~J1935+2154 has undergone another outburst since 2022 October 10. We present the results of searching for an optical/NIR counterpart of SGR~J1935+2154 before and during this outburst. No counterpart was detected at the magnetar's position in ${r'}$ and ${z'}$ bands, providing stringent upper limits of $r'\gtrsim 28.65$ and $z'\gtrsim 26.27$. 
Using archival X-ray data from NICER, we investigated the properties of the bursts and the spectral evolution of persistent emission.
The burst flux $F$ showed a power-law distribution of $N\propto F^{-0.76\pm0.10}$ for flux $\gtrsim 2.6\times 10^{-9}\rm{\ erg\ cm^{-2}\ s^{-1}}$, while the temperature and radius followed a lognormal distribution with $kT=1.63^{+0.73}_{-0.50}\ \rm{keV}$ and $R_{\rm bb}=4.35_{-1.35}^{+1.95}\ \rm{km}$, respectively. The persistent flux evolution experienced a quick decay and an enhancement  $\sim 27$ days after the BAT trigger.
Using the near-infrared (NIR) and X-ray emission, together with the GTC optical/NIR upper limits, we discussed the origin of the NIR emission from the magnetar based on the fallback disk model and magnetosphere model. We found that either model cannot be ruled out with currently available data. Further mid-infrared observations are needed to find out the mechanism for producing the NIR emission from SGR~J1935+2154.

\end{abstract}

\section{Introduction} \label{sec:intro}

Magnetars\ping{,} young rotating neutron stars with an \ping{extremely strong} magnetic field of $B\sim 10^{14}-10^{15}\ \rm{G}$, \ping{are traditionally falling into two subclasses}, the anomalous X-ray pulsars (AXPs) and soft gamma-ray repeaters (SGRs). They show long spin periods ($P\sim 1$--$12\ \rm{s}$) and \ping{high spin down derivative $\dot{P}\sim 10^{-11}$--$10^{-13}\ \rm s~s^{-1}$}.
\ping{The known Galactic magnetars have X-ray} persistent luminosity of $L_X\sim 10^{31}-10^{36}\ \rm{erg\ s^{-1}}$, \ping{with the high luminosity} believed to be powered by the decay of their enormous internal magnetic fields \citep{Duncan+1992ApJ}. Magnetars can undergo a wide \ping{range} of X-ray activities including short bursts, outbursts, giant flares, and quasi-periodic oscillations \citep[see][for a review]{Turolla+2015RPPh, Kaspi+2017ARA&A, Esposito+2021ASSL}. To date, the population of magnetars has grown to about 30 objects \citep{Olausen+2014ApJS}.

Although magnetars are \ping{usually discovered in the X-ray band}, some of them have detectable counterparts in optical/NIR wavelength \citep[e.g.,][]{Hulleman+2000Natur, Dhillon+2011, Chrimes+2022HST_counterpart}, while the origin of the optical/NIR emissions from magnetars is still under debate. The NIR spectrum energy distribution (SED) from magnetar 4U 0142+61 was found to be well fitted by a multi-temperature (700--1200 K) blackbody disk \citep{Wang+2006Natur}, suggesting a thermal origin from a debris disk passively heated by the X-ray emission from the magnetar, \ping{where the debris disk} was explained to form from the supernova fallback material. On the other hand, optical/NIR emission is also expected from nonthermal radiation in the magnetosphere.
\comm{Recently, a JWST observation of magnetar 4U 0142+61 shows the 1.1--14 $\mu m$ spectrum can be satisfactorily described by an absorbed power-law, suggesting the non-thermal origin \citep{Hare+2024ApJ_4U0142JWST}.}
The optical/NIR variability has been observed in some magnetars, \ping{but} whether it changes with the X-ray flux evolution is still unclear \citep[see][and the reference therein]{Esposito+2021ASSL}.  

SGR~J1935+2154 was discovered through a magnetar-like outburst in 2014 by the Burst Alert Telescope (BAT) on broad the Neil Geherls Swift Observatory \citep{Stamatikos+2014GCN}. The follow-up observations measured the spin-period of \ping{$P=3.24528(6)\rm\ {s}$} and spin-down rate of $\dot{P}=1.43\times 10^{-11}\ \rm{s\ s^{-1}}$, implying the source as a magnetar with inferred dipolar B-field of $B\approx 2.2\times 10^{14}\ \rm{G}$ and characteristic age $\tau_c \sim 3.6\ \rm{kyr}$ \citep{Israel+2016MNRAS}. This source has been sporadically active since it was discovered, \ping{and} has undergone several outbursts in 2015 February, 2016 May and June, and 2020 April, with multiple short bursts and X-ray burst forests detected \citep{Younes+2017ApJ,Borghese+2020ApJL,lin+2020ApJ}. After the outburst on 2020 April 27, a very bright FRB (FRB 200428) was detected from the direction of SGR~J1935+2154 by CHIME and STARE2 \citep{CHIME+2020Natur,Bochenek+2020Natur}. This was the first Galactic FRB event ever detected. The associated X-ray bursts were also detected by Insight-HXMT \citep{Li+2021NatAs}, confirming the association \ping{between} SGR~J1935+2154 and FRB 200428, \ping{and the magnetar as an origin of FRBs.}
SGR~J1935+2154 is among the few magnetars that emit radio pulses \citep{ZhuWeiWei+2023SciA}. Its radio pulsation phase is proposed to be associated with a change of X-ray properties \citep[e.g., X-ray hardening;][]{Wangpei+23} \comm{Additionally, some FRB-like radio pulsations appeared after a large spin-down glitch event during 2020 October, suggesting an association between these two events \citep{Younes+2023NatAs}.}

With the expansion of magnetar samples, the search for magnetars' optical/NIR counterparts has been ongoing for years. To date, there have been 19 out of about 30 magnetars \citep{Chrimes+2022MNRAS} with detectable NIR counterparts, a few magnetars have also been observed in optical (e.g., AXP 4U0142+61: \citealt{Hulleman+2000Natur}; SGR 0501+4516: \citealt{Dhillon+2011}; \hbox{AXP 1E 1048.1-5937}: \citealt{Wang+2008ApJ}). The NIR counterpart of SGR~J1935+2154 was detected with the Hubble space telescope (HST) in the F140W band \citep{Levan+2018ApJ}. The first three epochs from \cite{Levan+2018ApJ} showed a presumably associated variation between NIR and X-ray luminosity, while the later epoch in 2021 by \cite{Lyman+2022ApJ} indicated the NIR brightness not correlated with high energy activities. \ping{It is still unclear how the NIR emission is generated from SGR~J1935+2154, but finding emission in other bands (such as the optical band) can help to study this question.}

SGR~J1935+2154 underwent a new period of active state since 2022 October 10 \citep{Palm+2022ATel,2022ATelRoberts}. Since we observed the source on $r'$ and $z'$ bands with the Gran Telescopio Canarias (GTC) in 2022 August, we executed a follow-up observation after the outburst, in order to search for the optical/NIR counterpart of the magnetar before and during the outburst, thus providing constraints on the optical/NIR emission.

\ping{This paper presents} optical/NIR observations of SGR~J1935+2154 \ping{in} both quiescent state and active state with GTC telescope. \ping{We also analyzed the archival data from the Neutron star Interior Composition Explorer (NICER) to investigate the X-ray properties and evolution during and after the active state}. Section \ref{sec:obs} gives a description of the observations and data reduction. Section \ref{sec:result} presents the result of searching for the optical/NIR counterpart, the persistent X-ray emission, and the properties of the X-ray bursts. Section \ref{sec:discussion} discusses the fallback disk model and the magnetosphere origin scenario of optical/NIR emission. We summarize our work in Section \ref{sec:summary}.

\section{Observations and archival data} \label{sec:obs}

\begin{figure*}[htb]
    \centering
    \includegraphics[width=\textwidth]{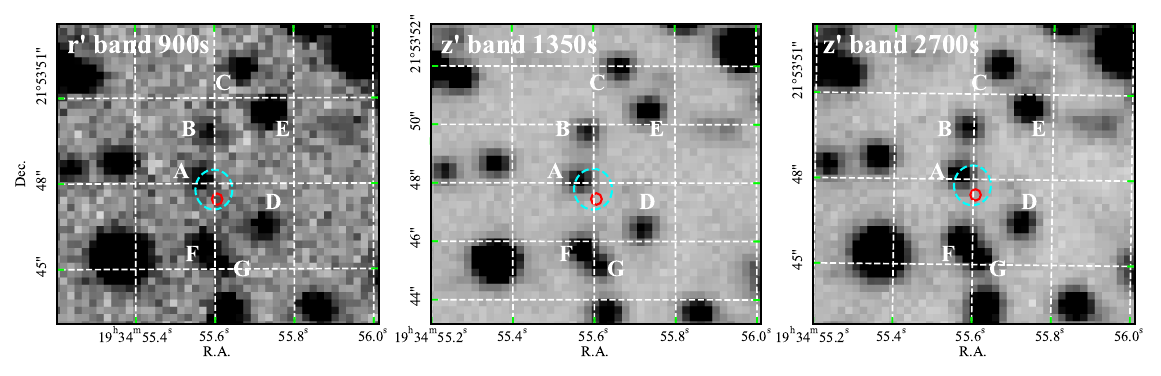}
    \caption{
    \ping{GTC optical/NIR images of SGR~J1935+2154. The three panels show the $r'$ band image with an exposure of 900 s, the 1st observation in ${z'}$ band with an exposure of 1350 s, and the 2nd observation on ${z'}$ band with an exposure of 2700 s, respectively. The cyan and red circles denote the 90\% uncertainties of the magnetar's position as determined by the Chandra X-ray observation and HST NIR observations respectively.}}
    \label{fig:location}
\end{figure*}

\subsection{GTC observations, Data reduction, and Calibration}

SGR~J1935+2154 \ping{returned} to a quiet state after the brief burst activities in 2022 January \citep{2022AtelMaan,2022ATelRoberts}. After 9 months of silence, the magnetar entered a new period of the active stage \ping{by showing} intense bursting activities in X-ray, gamma-ray and bright radio bursts \citep{2022ATelDong,2022ATelCHIME,2022ATelPalm}. The field of magnetar SGR~J1935+2154 was imaged twice with the Optical System for Imaging and low Resolution Integrated Spectroscopy \citep[OSIRIS;][]{Cepa+2000} mounted at the GTC. We performed an observation with the SDSS ${r'}$- and ${z'}$-band filters on August 18th, 2022 (GTCMULTIPLE4C-22A\_0032, PI: R.S. Ramríez) when the magnetar was still in a quiet state. On October 10th, 2022, X-ray bursts were detected by Swift/BAT \citep[i.e. 2022-10-10 01:39:09.0,][]{2022ATelPalm}, revealing \ping{that} SGR~J1935+2154 began active again. We subsequently proposed a Target of Opportunity (ToO) observation toward this source 
 and the observation was taken on 2022 October 29, with OSIRIS in ${z'}$ band.

The OSIRIS imager consists of 2 combined CCDs \ping{which combined to provide a $7'.8\times7'.8$ field of view (FoV) and pixel scale of $0''.254$ ($\rm 2\times2\ binning$)}. The observation was executed in the service mode under good observation conditions, with airmass variation under 4\% and seeing about $0.7''$. Considering the higher extinction effect in ${r'}$ band than ${z'}$ band \citep{Wang+2019ApJ} and the probable saturation of bright stars, the exposure time of individual images is \ping{set to} 100~s for ${r'}$ band and 50~s for ${z'}$ band. In addition, we used a dithering observation \ping{mode} to avoid the affection of bad pixels.

\begin{table}[h]
    \centering
    \caption{\textbf{
    \ping{Information of the GTC optical/NIR observations}}}
    \begin{tabular}{cccccc}
    \hline
    Date & \multirow{2}{*}{Filter} & Exposure time & \multirow{2}{*}{Airmass} & Seeing\\
    yyyy-mm-dd&&(s)&&(arcsec)\\
    \hline
    \multirow{2}{*}{2022-08-18}&${r'}$& $100\times9$ &1.01--1.04&0.7\\
    &${z'}$&$50\times27$&1.01--1.05&0.7\\
    \hline
    2022-10-29&${z'}$& $50\times54$ &1.10--1.29&0.7\\
    \hline  
    \end{tabular}
    \label{tab:log_obs}
\end{table}

We performed standard data reduction including bias subtraction, flat-fielding, and cosmic ray removal using the Image Reduction and Analysis Facility \citep[IRAF,][]{IRAF1,IRAF2}.\footnote{\ping{IRAF is distributed by the National Optical Astronomy Observatory,
which is operated by the Association of Universities for Research in Astronomy (AURA) under cooperative agreement with the National Science Foundation.}} In order to \ping{correct} the shifts of individual images, we chose a set of unsaturated stars and aligned the field of each observation set using the AstroImageJ \citep{astroImageJ} software. Finally, the combined images have mean seeings of $0.7''$, mean airmass of 1.03, 1.03, 1.18, and total exposure times of 900 s, 1350 s, and 2700 s, for ${r'}$, ${z'}$ (1st observation) and ${z'}$ (2nd observation) bands, respectively.

\begin{table*}[htb!]
    \centering
    \caption{\textbf{Astrometric solution and photometry results}}
    \begin{ThreePartTable}
    \begin{tabular}{cccccccc}
    \hline
    Date & \multirow{2}{*}{Filter} & $\Delta \rm R.A.$ \tnote{$\dagger$} &$ \Delta \rm\ decl. \tnote{$\dagger$}$&$1\sigma\rm\ uncertainties\ ('')$& \multirow{2}{*}{Upper limit\tnote{*}}& \multirow{2}{*}{Zero-points}\\
    yyyy-mm-dd&&(arcsec)&(arcsec)&R.A. decl.&&&\\
    \hline
    \multirow{2}{*}{2022-08-18}&${r'}$&0.022&0.032&0.201 0.202&$28.65$&$28.90\pm 0.01$\\
    &${z'}$&0.037&0.055&0.203 0.207&$26.03$&$27.89\pm 0.02$\\
    \hline
    2022-10-29&${z'}$&0.020&0.028&0.200 0.201&$26.27$&$27.84\pm 0.02$\\
    \hline  
    \end{tabular}
    \begin{tablenotes}
        \item[$\dagger$] The rms uncertainties of the astrometric fit for the combined images.
        \item[*] $3\sigma $ upper limits.
    \end{tablenotes}
    \label{tab:astrometric and photometric}
    
    \end{ThreePartTable}
\end{table*}

For astrometric solutions for these 3 pulsar fields, we chose 3 sets of bright and unsaturated stars from 2MASS point source catalog as Web Coverage Service (WCS) reference stars \citep{2mass+2003}, numbering 54, 50, and 51 for ${r'}$, ${z'}$ (1st observation) and ${z'}$ (2nd observation) bands, respectively. The coordinates were measured using the IRAF task \textit{imexamine} on the combined images. To correct the geometry distortion of OSIRIS fields \ping{as} described in the OSIRIS User Manual\footnote{\url{http://www.gtc.iac.es/instruments/osiris/media/OSIRIS-USER-MANUAL_v3_1.pdf}}, we used the \textit{CCMAP} routine in IRAF for a precise astrometric solution. The \ping{$1\sigma$} uncertainties of the images are presented in Table~\ref{tab:astrometric and photometric}. Accounting for the nominal catalog uncertainty of $\approx0.2''$ , the resulting conservative $1\sigma$ referencing uncertainties of the astrometric fitting are also presented in Table~\ref{tab:astrometric and photometric}, which is consistent with the nominal catalog uncertainty.

The photometric calibration was carried out with PG 1528+062B and SA 111-1925 Sloan standard stars \citep{Smith+2002AJ} observed the same night as our two observations. We used the measured instrument magnitudes of these standard stars and the mean OSIRIS atmosphere extinction coefficients $k_{r'}=0.07\pm0.01$, $k_{z'}=0.03\pm0.02$ from OSIRIS broad imaging calibration guide\footnote{\url{http://www.gtc.iac.es/instruments/osiris/media/CUPS_BBpaper.pdf}}. The resulting zero-points of our three observations are presented in Table~\ref{tab:astrometric and photometric}.

\subsection{NICER archival data, filtering and calibration}
\label{sec: 2.2}
NICER is a non-imaging X-ray timing and spectral soft X-ray instrument onboard the International Space Station (ISS) to study neutron stars in the 0.2--12 keV band \citep{Gendreau+2016}. NICER provides a large \comm{collecting} area peaking at $1900\rm \ cm^2$ at 1.5~keV and \ping{a} timing resolution $<300\rm\ ns$. NICER conducted extensive observations of SGR~J1935+2154 since the outburst was detected from this magnetar \citep{Palm+2022ATel,2022ATelRoberts}. We retrieved the archival NICER monitoring data for about eleven months, starting on 2022 October 12 \ping{(MJD 59864)}, and ending on 2023 September 12 \ping{(MJD 60199)}. The observational information of the final used data is presented in Table~\ref{tab: persistent spec}.

To process the data, we used \texttt{HEASoft v6.32.1} and the NICER specific package \texttt{NICERDAS v011a} with Calibration Database (\texttt{CALDB vXTI20221001}). We applied standard filtering criteria with \textit{nicerl2} task:
 
 \begin{enumerate}[1)]
     \item Offset from the target $< 0.015\rm \ deg$.
     \item Constrain of elevation $>15\rm\ deg$ and $>30\rm\ deg$ in case of bright earth.
     \item Exclude the NICER-specific definition of the South Atlantic Anomaly (SAA) region stored in CALDB.
 \end{enumerate}

Moreover, we excluded detectors with \textit{DET\_ID} 14, 34, and 54 because these are known to be more sensitive to optical loading than others. Considering \ping{that} the ``noise ringers" \footnote{{\url{https://heasarc.gsfc.nasa.gov/docs/nicer/analysis_threads/noise-ringers/}}} which may occur in the soft energy range (0.2--0.6 keV) and \ping{thus results in an incorrect} hydrogen column density \ping{in the spectral fit of the source}, we extracted \ping{spectra only in} the 1--7 keV for our analysis. To filter the high background regions and the non-X-ray flares as described in the NICER Data Analysis Threads\footnote{\url{https://heasarc.gsfc.nasa.gov/docs/nicer/analysis_threads/}}, we applied the constraints on the cut-off rigidity (\comm{i.e. COR\_SAX}) to the range \ping{\comm{COR\_SAX}$>$1.5}, also the overshoot count rate was limited to 0--1 for a strict filter. We compared the light curve of 1--5 keV to the 13--15 keV (which is considered from the high-energy background), any simultaneous flaring intervals \ping{occured in these two bands} were finally eliminated in our level 2 data. \rep{We used the Chandra position \citep[R.A.=19:34:55.5978, decl.=+21:53:47.7684,][]{Israel+2016MNRAS} and the JPL planetary ephemeris DE 200 for the barycenter correction of the data.} Finally, we used \textit{nicerl3-lc} to extract lightcurves. The source spectra were extracted from the level 2 data utilizing \textit{nicerl3-spect} pipeline, \ping{and} the background was estimated using the nibackgen3C50 method \citep{Remillard+2022AJ3C50}. The spectral analysis was performed with XSPEC version 12.13.1, we chose W-statistic (c-stat in XSPEC) to estimate the model parameters and calculate the error.


\section{Results} \label{sec:result}

\subsection{Searching for the optical/NIR counterpart}

The X-ray position of SGR~J1935+2154 is R.A.=19:34:55.5978, decl.=+21:53:47.7864, with a 90\% uncertainty of $0.7''$ \citep{Israel+2016MNRAS}. Previous HST observation has identified the NIR counterpart of this SGR~J1935+2154 \citep{Levan+2018ApJ} at R.A.19:34:55.606 decl.=21:53:47.45 ($\pm 0.2''$).

\ping{Figure~\ref{fig:location} shows the GTC images of the magnetar field}. The X-ray position from the previous Chandra observation is marked \ping{with} \ping{a} cyan circle. The red circle shows the position and the  $1\ \sigma$ position uncertainty ($\sim 0.2''$) of the HST NIR counterpart. The NIR counterpart of SGR~J1935+2154 is located in the wings of source A  as shown in Figure~\ref{fig:location}, but no significant point-like object is detected within the error circle in our GTC images. Moreover, comparing with the HST NIR counterpart \citep{Levan+2018ApJ}, we confirmed that the NIR counterpart is not detected in our GTC observations.

\begin{figure*}[htb]
    \centering
    \includegraphics[width=\textwidth]{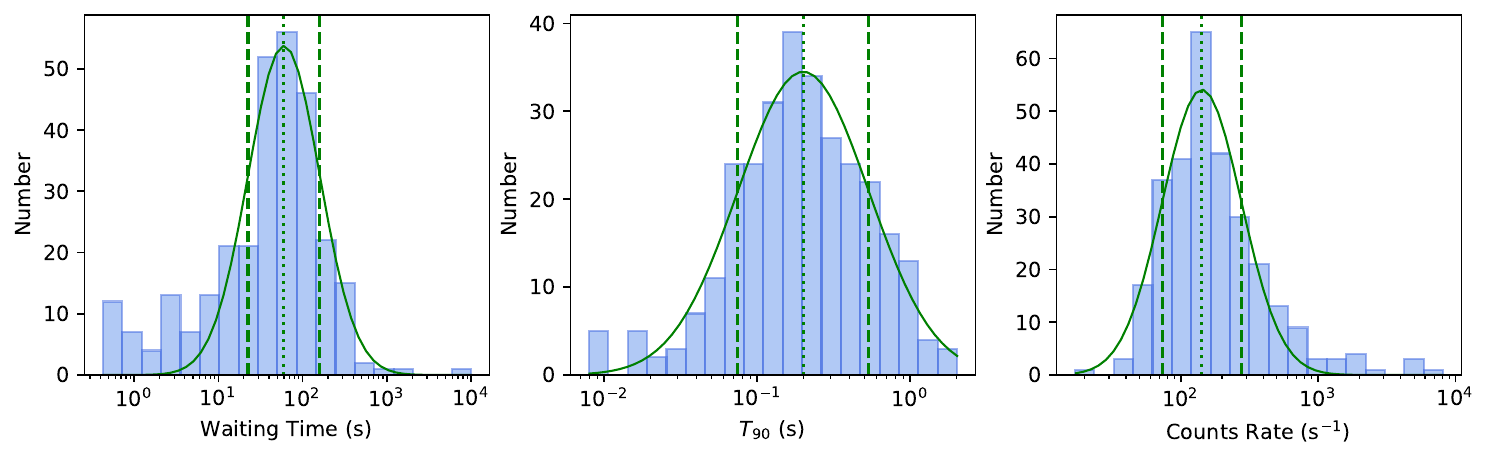}
    \caption{Left panel: distribution of burst waiting time fit with the lognormal function with a mean of $59.94\ \rm{s}$ and $1 \sigma$ range of 22.48--159.80 s. Middle panel: distribution of $T_{\rm 90}$ and the best-fit lognormal function with a mean of $0.20\rm\ s$ and $1\sigma$ range of 0.07--0.53 s. Right panel: distribution of burst counts rate and the best-fit lognormal function, with a mean of $142.41\ \rm s^{-1}$ and $1\sigma$ range of 73.36--276.47 $\rm{s^{-1}}$. The green line shows the best-fit lognormal distribution and the vertical dashed lines show the mean value and $1\ \sigma$ range of the parameter.}
    \label{fig:bur_time}
\end{figure*}

Since \ping{the GTC optical/NIR observations did not reveal a counterpart} in the SGR 1934+2154 \ping{field}, we can only set upper limits on the \ping{optical/NIR emission of the} magnetar. \ping{From} the source-free regions of the three images, \ping{we obtained} the $3\sigma$ deviation upper limits of $r'\gtrsim 28.87$, $z'\gtrsim 26.08 \rm (1st)$ and $z'\gtrsim 26.35\rm (2nd)$, respectively. 
We chose ten unsaturated bright and well-imaged stars in the \ping{vicinity} of the magnetar for measuring the point spread function (PSF). \ping{Using the IRAF \textit{imexamine} task, the FWHM of these sources for the three images are determined to be \ping{$2\farcs{75}$, $2\farcs{72}$, and $2\farcs{83}$}, respectively.} \ping{The $3\sigma\ \rm radius =\frac{3}{2\sqrt{2ln(2)}}\rm\ FWHM$} is significantly larger than the distance between the peak of source A and the \ping{HST NIR counterpart of SGR J1935+2154} \ping{($\sim 0.6 ''$), suggesting that the location of SGR~J1935+2154 could be dramatically altered due to the influence of} the PSF of the bright source A in Figure~\ref{fig:location}.
Therefore, a PSF photometry is needed to robustly estimate the upper limits of SGR~J1935+2154. \rep{We used the chosen stars above }to build a normalized PSF model, using the Photutils task \textit{EPSFBuilder}, with a $3\sigma$ norm radius, then the PSF model was used to fit the bright source in a $\sim 8''\times 4''$ region, containing the marked sources in Figure~\ref{fig:location}. The Photutils tasks \textit{IRAFStarFinder} and \emph{IterativelySubtractedPSFPhotometry} were used to detect the sources and fit the PSF \ping{in} the background-subtracted image. 
\ping{All 7 nearby bright sources (see Fig \ref{fig:location} A--G) in the field were fitted and subtracted within 2 iterations and resulted in a clean and smooth residual image}. \ping{Using the above approach, we did not find any counterpart of SGR~J1935+2154 in the residual images.} The $3\sigma$ upper limits at the magnetar position \ping{based on} the PSF-subtracted images are \ping{$r'\sim 28.65$, $z'\sim 26.03 \rm ~(1st)$ and $z'\sim 26.27\rm ~(2nd)$}, respectively.

\subsection{X-ray properties of the 2022 outburst}

\rep{The magnetospheric model of NIR/optical counterpart predicts a correlation with the X-ray emission\citep{Beloborodov+2007ApJ}, but it's still unclear whether the changes of NIR/optical flux trace the X-ray activity. Since we have executed optical/NIR observation during the active stage \ping{in 2022, we also analyzed the X-ray data to understand the high-energy behaviors.} This section presents the NICER view of SGR~J1935+2154's activity, including the properties of the bursts and the persistent emission.}

\subsubsection{Burst identification and properties}
\label{subsec:bursts}

We performed a search for short X-ray bursts based on all available data from NICER after the burst alert \citep[MJD 59862.0688; 2022-10-10T01:39:09 UTC;][]{2022ATelPalm}, starting at MJD 59864.8534 (2022-10-12T20:28:52 UTC). 
We executed the same Possionian procedure \ping{for our light curves as used in \cite{Gavriil+2004ApJ...607..959G} and \cite{Younes+2020ApJL}}. First, we obtained the light curves of each observation \ping{with} different resolutions, $\Delta \tau$. We divided the good time intervals (GTIs) into multiple time intervals with 100 s duration, $\Delta T$, then resulting \ping{number of bins} $N=\Delta T/\Delta \tau \rm{\ bins\ interval^{-1}}$, \ping{and} the counts number in each bin \ping{$i$ of} $n_i$. Using the Poisson probability $P_{i}=(\lambda^{n_i}exp(-\lambda))/n_i!$, any time bin with a probability $<10^{-2}/N$ was flagged as part of a candidate burst. To ensure the completeness of the searching procedure and prevent missing the faint tail of strong bursts, the same procedure was repeated \ping{for} lightcurves with resolutions of 8, 16, 32, 64, 128, 256, and 512~ms. We also repeated the same algorithm \ping{with} different time intervals, including 20, 50, and 200~s, and we found the choice of $\Delta T$ has a weak effect on the result, consistent with \ping{that found in} \citep{Younes+2020ApJL}.

\rep{After flagging all candidate time bins under different time resolutions, we assembled these time bins. The criterion of several time bins to be grouped into a single burst is that the emission \ping{does not} drops to the background level between two neighboring time bins. In that case, we obtained the ${T_{\rm start}}$ and ${T_{\rm stop}}$ of each burst, waiting time between neighboring bursts and the ${T}_{\rm 90}$ (the 5\%--95\% interval of accumulated counts of each burst.}

\rep{We identified 371 bursts within 16 epochs as shown in Table~\ref{tab: burst spec}. We fit the waiting time, $T_{\rm 90}$, and counts of each burst with a lognormal distribution, as shown in Figure~\ref{fig:bur_time}. The fitting results show the waiting time with a mean of $59.9\ {\rm s}$ and a $1 \sigma$ range of $22.45\text{--}159.8\ \rm s$, significantly larger than the measurement in 2020's burst storm \citep[$-0.5\pm0.1$,][]{Younes+2020ApJL}, but similar with the measurement of magnetar 1E 2259+586's outburst in 2002 June \citep[$0.7\pm 0.1$][]{Gavriil+2004ApJ...607..959G}. The fitting for ${T}_{\rm 90}$ and counts of bursts show ${T}_{\rm 90}=0.20_{-0.12}^{+0.33}\ \rm{s}$ and Counts Rate=$142.41_{-69.05}^{+134.06}\rm \ s^{-1}$.}

\subsubsection{Burst spectroscopy}
\label{subsec:3.2.2}
We extracted the GTI file of each burst identified by our Possionian procedure, then used the \textit{nicerl3-spec} task with \textit{bkgmodeltype=none} option to extract a source spectrum without background estimation. \comm{\sout{using the \textit{nibackgen3C50} tool.}} The background of each burst is estimated from an 80-s long interval centered on each burst peak and eliminated other bursts during this interval.
For the spectra fitting, we adopted the \textit{tbabs} model to account for the line-of-sight interstellar absorption, along with the elemental abundance from \citep{Wilms+2000} and the photoelectric cross sections of \citep{Verner+1996ApJ}. We fit the 1--7 keV spectra of all bursts using an absorbed BB model with the hydrogen column density fixed to $N_{\rm H}=2.3\times10^{22}\rm{\ cm^{-2}}$, which is derived by \cite{Coti+2018MNRAS}.  This BB model could fit most bursts well, except for some with too few counts for any spectrum. The fitted results of the bursts are tabulated in Table~\ref{tab: burst spec}.

\begin{figure*}[htb]
    \centering
    \includegraphics[width=\textwidth]{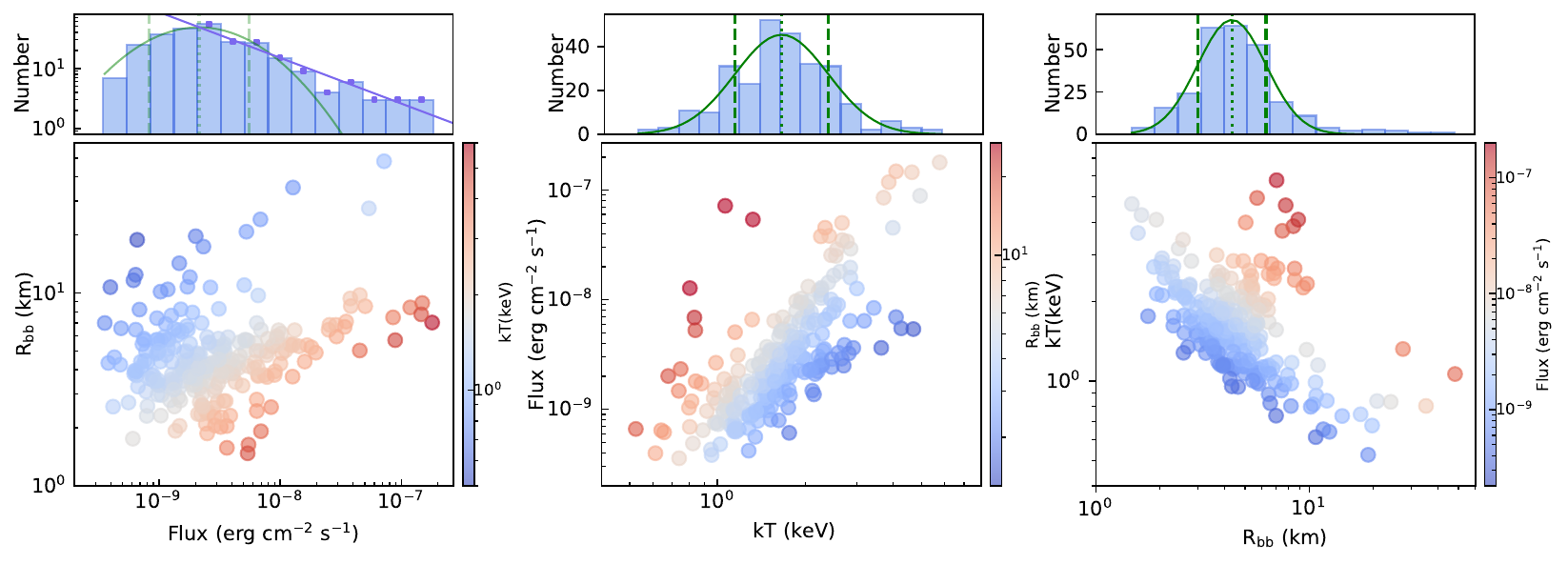}
    \caption{\ping{Distribution of spectral fit results of bursts from SGR~J1935+2154. Each data point in the lower panels represents one burst.} Left panel: BB radius vs.\ flux in 0.3-10 keV, with the color representing the BB temperature. The upper panel shows the distribution of the flux, which can not be fitted by a lognormal function (green), 
    but follows a power-law with $N\propto F^{-0.76\pm{0.10}}$ (purple dots and solid line) for flux $\gtrsim 2.6\times 10^{-9}\rm{\ erg\ cm^{-2}\ s^{-1}}$. Middle panel: flux vs.\ BB temperature, with the color representing the BB radius. The upper panel shows the distribution of BB temperature fitted with a lognormal function with a mean of $1.63\rm{\ keV}$ and $\rm{ 1\sigma\ range\ of\ \text{1.13--2.36}\ keV }$. Right panel: BB temperature vs.\ BB radius, with the color representing the flux. The upper panel shows the distribution of BB radius fitted with a lognormal function with a mean of $4.35\ \rm{km}$ and  1$\sigma$ range of 3.00--6.30 km). The vertical dashed lines show the mean value and $1 \sigma$ range of the parameter.}
    \label{fig:bur_statistic}
\end{figure*}

Figure~\ref{fig:bur_statistic} shows the distributions of the \ping{best-fit} fluxes, $kT$, and $R_{\rm bb}$ of all \ping{detected} bursts. 
The burst flux distribution cannot be fitted with a lognormal function, but follows a power-law with $N\propto F^{-0.76\pm0.10}$ for $\rm{flux}\ F\gtrsim 2.6\times10^{-9}\ erg\ cm^{-2}\ s^{-1}$, corresponding to $dN/dF \approx F^{-1.76\pm0.10}$. 
The 1-D distribution of $kT$ and $R_{\rm bb}$ followed a lognormal distribution with $kT=1.63_{-0.50}^{+0.73}\ \rm{keV}$  \citep[similar to the 2020 outbursts,][]{Younes+2020ApJL} and $R_{\rm bb}=4.35_{-1.35}^{+1.95}\ \rm{km}$, respectively. 


\subsubsection{\ping{Spectral analysis of the long-term persistent emission}}
\label{subsec:persistent}

We fit the 1.0--5.0 keV spectra of all available observations \rep{as mentioned in \ref{sec: 2.2}, after excluding all identified bursts in \ref{subsec:bursts}}. 
The persistent spectra were grouped with at least 50 counts in each bin and the $\chi^{2}$ statistics were adopted. We fitted the spectra with an absorbed two-blackbody (2BB) model and a blackbody plus a power-law (BB+PL) model, respectively. 
Following the previous studies \citep{Coti+2018MNRAS,Borghese+2020ApJL}, the hydrogen column density is fixed to \ping{$N_{\rm H}=2.3\times10^{22}\ \rm cm^{-2}$}. Because of the \ping{narrow energy range}, the photon index could not be well constrained, hence \ping{we fixed} the photon index to $\Gamma=1.2$ \ping{as did in previous studies \citep{Borghese+2020ApJL}}. 
\comm{Most of these spectra could be better described with an absorbed BB+PL model (see Figure~\ref{fig: persistent spectral} for the spectra), consistent with the previous studies \citep{Borghese+2022MNRAS, Younes+2017ApJ}.}
The best fitting parameters with their $90\%$ confidence level are listed in Table~\ref{tab: persistent spec}.

\begin{figure*}[htb]
\centering
  \includegraphics[scale=0.8]{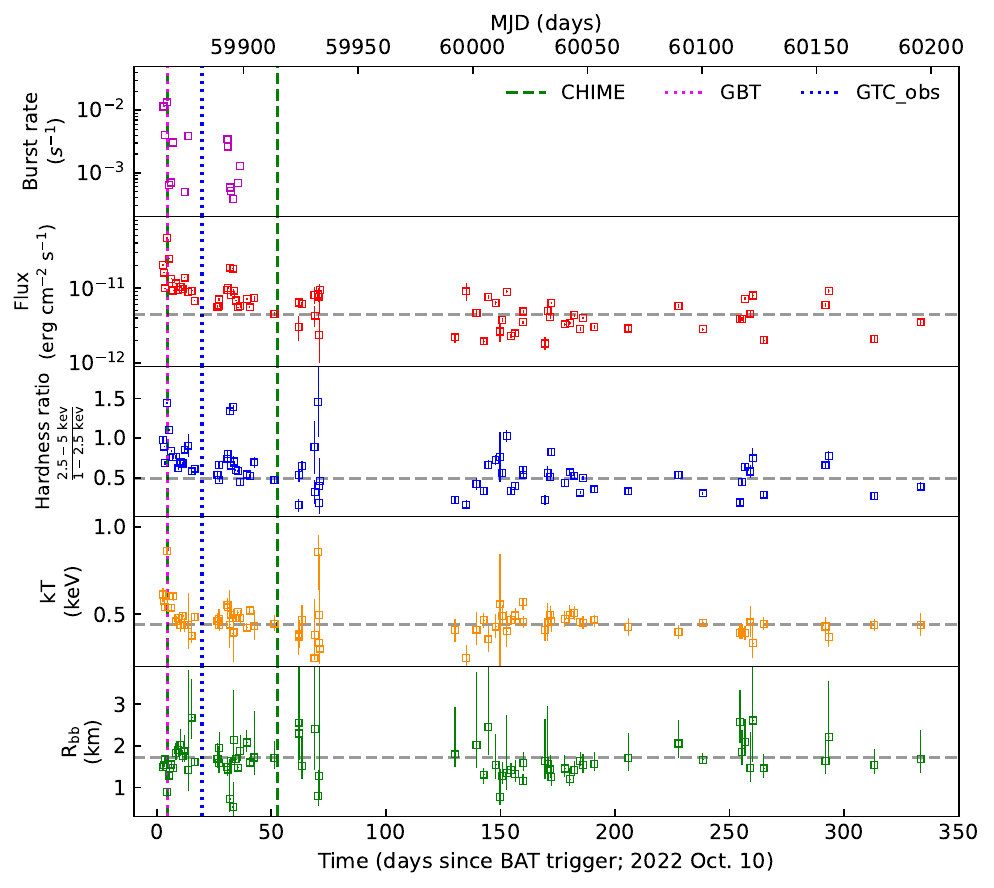}
  \caption{\ping{Evolution of bursts (upper panel; during the outburst stage) and persistent emission of SGR~J1935+2154 }
  \rep{The flux is measured in 0.3--10 $\rm keV$ range}. \ping{The blackbody temperature $\rm kT$ and radius $R_{\rm bb}$ are obtained from the spectral fitting in the 1--5 keV range.} The green and magenta vertical lines mark the radio bursts events alert by CHIME and GBT \citep{2022ATelDong,2022AtelMaan,2022ATelCHIME}, respectively. The blue dotted line between them marks our second GTC observation. The horizontal gray dashed line shows the average of the values after MJD 60021.9423 (2023-3-18T22:36:56 UTC), representing a quiescent level $\sim 150$ days after the burst onset.}
  \label{fig:persistent para}
\end{figure*}

\rep{In Figure~\ref{fig:persistent para}, we show the evolution of the burst rate, X-ray flux (0.3--10 $\rm keV$), hardness ratio (2.5--5 keV/1--2.5 keV), blackbody temperature ($kT$), and the radius of the hot emission region estimated from the BB comment ($R_{\rm bb}$).} The persistent flux reached its peak value of $4.67\times 10^{-11}~{\rm  erg\ cm^{-2}\ s^{-1}}$ at around the first radio burst on MJD 59866.8 \citep{2022ATelCHIME, 2022AtelMaan}. The emission quickly decays and experiences an enhancement after $\sim27.1$ days, while the previous outbursts of this magnetar only showed (double) exponential decay\citep{Younes+2017ApJ,Younes+2020ApJL}. \rep{The highest burst rate is $0.013\ \rm s^{-1}$, consistent with the measurement in 2022 \citep{Younes+2022ATel}, but significantly smaller than that of 2020's burst storm \citep[$>0.2\ \rm s^{-1}$,][]{Younes+2020ApJL} \ping{during which period the FRB was detected}.} The evolution of the hardness ratio and the flux appear to be correlated. \ping{The} temperature also appears to show a strong increase associated with flux \rep{before MJD 55950}, \ping{while} $R_{\rm bb}$ shows no apparent \ping{correlation with the other three parameters}. The magenta and green dashed line \ping{marks} the radio burst detected by GBT and CHIME, respectively \citep{2022ATelDong,2022ATelPalm,2022ATelCHIME}. The blue dotted line represents our second GTC observation, which \ping{was performed at around} the end of the first active stage's declining trend, before the second enhancement of persistent flux.

\section{Discussion} \label{sec:discussion}

The mechanism of NIR/optical emission of magnetars is still under debate. One of the explanations is the thermal radiation from the fallback disk that is heated by the X-ray emission from magnetars \citep{Perna+2000ApJ}, and the alternative one is non-thermal radiation from the magnetosphere \citep{Beloborodov+2007ApJ}. \ping{These two mechanisms predict different broad-band spectra of magnetars. Hereafter, we discuss the predicted spectral energy distribution (SED) of the two scenarios and compare them with the NIR/optical/X-ray observations.}

\subsection{Fallback disk scenario}
\label{subsec:fallback}
After the supernova explosion \ping{of a massive star}, some \ping{ejecta} material would fall back with the reverse shock and gain angular momentum during the falling, finally forming a debris fallback disk around the neutron star \citep{Alpar+2001ApJ,Chatterjee+2000}. This fallback disk can be passively illuminated by \ping{the bright} X-ray emission from neutron star and observed in the IR band \citep{Wang+2006Natur}. 

In the X-ray-illuminated disk model, the blackbody temperature of the disk is \cite{Vrtilek+1990A&A}:
\begin{equation}
T(r)\approx2088~{\rm K}~(1-\eta)^\frac{2}{7} (\frac{L_{\rm X}}{2\times 10^{34}\rm{\ erg\ s^{-1}}})^{\frac{2}{7}} (\frac{r}{R_\odot})^{-\frac{3}{7}}
\end{equation}
where $\eta$ is the X-ray albedo, $L_{\rm X}$ is the X-ray luminosity of the neutron star, and $r$ is the radius of the emission region on the disk.

The emissivity of the disk with an inner radius \ping{$r_{\rm in}$} and outer radius \ping{$r_{\rm out}$} is given by \citep{Frank+2002apa..book.....F}: \rep{
\begin{equation}\label{eq:fe}
F_E=(\frac{r_{\rm in}}{D}^2)\frac{4{\pi}E^3cos(i)}{h^3c^2}\int_1^{x_{out}}\frac{x}{e^{\frac{E}{k\ T(r)}}-1}dx\ (x=r/r_{\rm in})
\end{equation} }
\ping{where $i$ is the inclination angle $i$ of the disk (0 degrees for a face-on disk, 90 degrees for an edge-on disk), $E$ is the photon energy, $D$ is the distance from observer to the disk, $c$ is the speed of light, $h$ is the Planck constant, and $k$ is the Boltzmann constant.}

\rep{Since the X-ray luminosity of SGR~J1935+2154 is highly variable, we took a few representative values to predict the fallback disk emission. From the nearest NICER observation before our second GTC observation (OBS5576010111 on 2022-10-26), we obtained a persistent luminosity $L_{\rm X1}=3.54\times{10^{34}}\rm{\ erg\ s^{-1}}$ in 0.3--10~keV for the magnetar at a distance of $6.6\rm{\ kpc}$ \citep{Zhou+2020ApJ}. We also considered the averaged and maximum persistent X-ray luminosity  $\bar{L_{\rm X}}=7.65\times{10^{34}}\rm{\ erg\ s^{-1}}$ and $L_{\rm Xmax}=2.43\times{10^{35}}\rm{\ erg\ s^{-1}}$, respectively, from the onset of the bursts till the second GTC observation.}

The currently available data for comparison with the fallback disk model include  HST F140W NIR values and our GTC $z'$- and $r'$- band upper limits.
For extinction correction, we adopted $A_\lambda\ /A_{V}=0.1939\pm 0.0048, 0.458\pm0.003, 0.848\pm0.006$ for HST F140W, SDSS $z$, and SDSS $r$ band, respectively \citep{Wang+2019ApJ}.
The extinction at the $V$ band was calculated by using 
the empirical relation between interstellar extinction and the hydrogen column density: $N_{\rm H}=(2.87\pm0.12)\times10^{21} A_V\ \rm{ cm^{-2}}$ \citep{Foight+2016ApJ}, where $N_H=2.3\times10^{22}\ \rm{cm^{-2}}$ \citep{Coti+2018MNRAS}. The extinction values of the magnetar in these three bands are obtained as $A_{\rm F140W}=1.554, A_{z'}=3.670, A_{r'}=6.796$, respectively.

\begin{figure}[htb]
    \centering
    \includegraphics[width=0.5\textwidth]{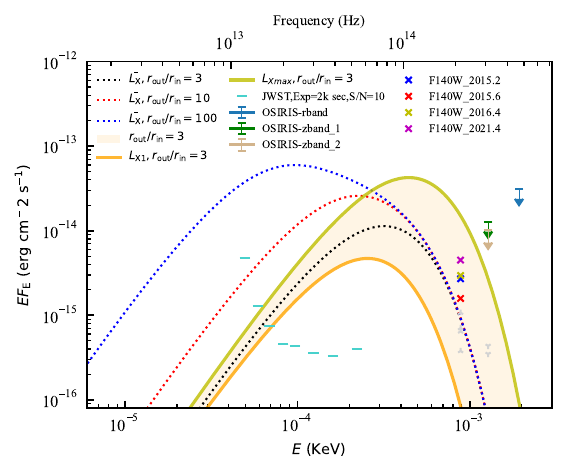}
    \caption{
    \ping{The SEDs of the disk-blackbody models in comparison with the data in IR and optical/NIR bands. 
    The dotted lines are the SEDs for $r_{\rm in}=3R_{\odot}$, $r_{\rm out}/r_{\rm in}=3$--100, and magnetar flux of $\bar{L_{\rm X}}$. The orange-shaded region shows the SED for $L_{\rm X}$ between $L_{\rm X1}$ and $L_{\rm Xmax}$.
    The NIR fluxes from HST F140W observations \citep{Levan+2018ApJ, Lyman+2022ApJ} and the upper limits in the optical/NIR bands with our GTC observation
    are denoted using the ``$\times$'' signs and arrows, respectively. The color and grey data points correspond to values with and without the extinction correction, respectively.  The turquoise ``-'' signs show the predicted $10\ \sigma$ sensitivity of a 2~ks JWST MIRI observation.}}
    \label{fig:fallback_SED}
\end{figure}

The SED of the fallback disk is shown in Figure~\ref{fig:fallback_SED}, \ping{in which} we considered a disk with the inclination angle of $i=60^{\circ}$. 
\cite{Wang+2006Natur} found that the IR and optical emission of the magnetar 4U~0142+161 can be describe by a debris disk with $r_{\rm in}=2.9R_{\odot}$ and $r_{\rm out}=9.7R_{\odot}$. Here we assume a similar value of $r_{\rm in}$ and $r_{\rm out}/r_{\rm in}=3$ for SGR~J1935+2154, although the real values can be very different.
The orange shaded region in Figure~\ref{fig:fallback_SED} shows the predicted disk spectra for the magnetar with an X-ray luminosity between $L_{\rm X1}$ and $L_{\rm Xmax}$. We note that the spectra are highly sensitive to the X-ray albedo $\eta$, which was assumed to be $\eta=0.5$ in \cite{Vrtilek+1990A&A}, and was fitted as $\eta=0.97$ in \cite{Wang+2006Natur}. These two values are found to over-predict and under-predict the observed NIR flux. Assuming $\eta=0.8$, we found that the spectra can cover the observed NIR data (``$\times$'' signs with color).

Although our GTC observation has already given the deepest upper limits, the available optical/NIR points are insufficient to constrain the disk parameters. 
More data points in IR and optical bands are needed to fit the disk black body model and to obtain the disk parameters. However, Figure~\ref{fig:fallback_SED} shows that a few groups of disk values may explain the NIR/optical observations, and suggest that the fallback disk model cannot be ruled out to explain the NIR and optical emission from this magnetar.

\comm{MIRI} observations will be useful to test the fallback disk model, since the disk SED peaks at the MIR bands (see in Figure~\ref{fig:fallback_SED}).
The turquoise ``-'' signs give the predicted $10\ \sigma$ sensitivity of the mid-infrared instrument (MIRI) on James Webb Space Telescope (JWST) with a short exposure of 2~ks. It shows that a JWST observation is able to give a good constraint on the model.

\begin{figure}[htb]
    \centering
    \includegraphics[width=0.5\textwidth]{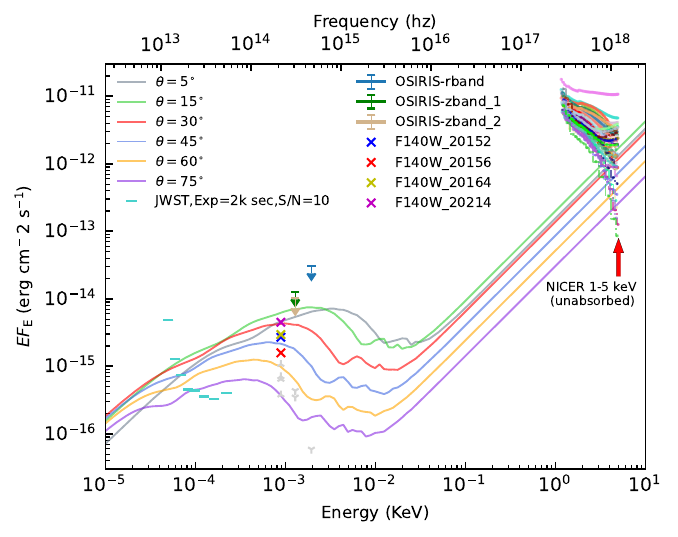}
    \caption{
    Coherent CR spectra from the bunched particles \citep{Zane+2011ASSP} and the data from NIR, optical, and X-ray bands.
    The different colors of the solid lines show different viewing angles. 
    The NIR flux from HST F140W observations and upper limits in our GTC observations are denoted using the ``$\times$'' signs and arrows, respectively. The turquoise ``-'' signs show the predicted $10\ \sigma$ sensitivity of a 2~ks JWST MIRI observation. The unabsorbed 1--5 keV X-ray spectra from NICER is also in the plot.
    }
    \label{fig:curvature}
\end{figure}

\subsection{Magnetosphere origin scenario}

In addition to the fallback disk model, \comm{another possibility is the nonthermal origin of the NIR/optical emission from magnetars, which has been favored by recent IR observation on magnetar 4U 0142+61 with JWST \citep{Hare+2024ApJ_4U0142JWST}. Here we discuss} the coherent curvature radiation from the pair-dominated region in the inner magnetosphere \citep{Zane+2011ASSP,Beloborodov+2007ApJ}. In this scenario, the high-energy photons produced by resonant cyclotron scattering (RCS) could produce the $e^{\pm}$ pairs, which are accelerated by the electric field and ultimately emit through curvature radiation. 
A supporting case for this magnetospheric origin is the detection of the optical pulsations from magnetars (4U 0142+61, 1E 1048$-$5937, and SGR J0501+4516; see \citealt{Dhillon+2011} and the reference therein), in which the similar pulse profiles between the optical and X-ray bands favors the nonthermal origin of the optical emission.

According to the models by \cite{Zane+2011ASSP}, the coherent curvature radiation spectrum highly depends on the angles between the magnetic axis and the line-of-sight (viewing angle, $\theta_{obs}$), with
a higher flux for a small $\theta_{obs}$ (see Figure~\ref{fig:curvature}). 
This model suggests that the coherent process of the bunched particles may be a possible channel to produce the IR/optical emission from some magnetars. 
As shown in Figure~\ref{fig:curvature}, the curvature radiation might explain the NIR emission and optical upper limits in SGR~J1935+2154. We note that this nonthermal radiation is insufficient to account for the observed soft X-ray emission. This agrees with the results that soft X-ray spectra can be described with black body radiation. 
In Figure~\ref{fig:curvature}, we also plotted the sensitivity of a 2~ks JWST/MIRI (the turquoise ``-'' signs) observation to show that JWST  will help to distinguish between the magnetosphere model and the fallback disk model (see also Figure~\ref{fig:fallback_SED}).

\comm{\subsection{Comparison to the previous outburst}}
\comm{Here we compare the outburst in 2022 with the previous outburst that occured in 2020. 
In 2022, the burst flux distribution follows a power-law with $N\propto F^{-0.76\pm0.10}$ for $\rm{flux}\ F\gtrsim 2.6\times10^{-9}\ erg\ cm^{-2}\ s^{-1}$, corresponding to $dN/dF \approx F^{-1.76\pm0.10}$. It is steeper than the distribution of burst storm of SGR~J1935+2154 in 2020 \citep[$-0.5\pm0.1$,][]{Younes+2020ApJL}, but comparable to the flux distribution of X-ray bursts from some other magnetars (e.g., 1E 2259+586 by \citealt{Gavriil+2004ApJ...607..959G}, SGR J1550-5418 by \citealt{Lin+2013ApJ}). This flux distribution is also similar to the Gutenberg-Richter distribution for earthquakes as mentioned in \cite{Cheng+1996Natur}. Moreover, this power-law distribution is characteristic of a self-organized criticality system, which is very common in a variety of natural dynamical phenomena, including earthquakes, wildfires, solar flares, etc \citep{Aschwanden+2013}.}

\comm{
As shown in Figure~\ref{fig:bur_statistic}, the noticeable positive correlation between $R_{\rm bb}$ and flux, $kT$ and flux indicates that stronger bursts with higher flux have higher temperatures, consistent with the outburst in 2020 \citep{Younes+2020ApJL}. The right panel of Figure~\ref{fig:bur_statistic} shows an anti-correlation between $kT$ and $R_{\rm bb}$, which is similar to the previous study by \cite{lin+2020ApJ}, indicating that hotter bursts are from smaller regions.}

\section{Summary} 
\label{sec:summary}

We performed two optical/NIR observations of the SGR~J1935+2154 before and during the outbursts of Oct.\ 2022, respectively, with the GTC OSIRIS in $r'$ and $z'$ bands, 
to search for the optical/NIR counterpart of the magnetar SGR~J1935+2154. No counterpart was significantly detected \ping{at the magnetar's position}, setting the deepest upper limit of \ping{the optical/NIR flux of} the source: $r'\gtrsim 28.65$, $z'_1\gtrsim 26.03$, and $z'_2\gtrsim 26.27$. 

We also analyzed the X-ray data during and after the 2022 outburst from NICER, with 76 available observations \ping{spanning} over 290 days. The persistent \ping{emission} shows dramatic variation \ping{of spectral properties} until $\sim 50$ days after the Swift/BAT trigger (Figure~\ref{fig:persistent para}). The flux did not monotonously decay with time, but enhanced around 27 days after the onset of the outburst. 
This suggests that the magnetar was still active during our second GTC observation.
By analyzing the burst properties, we found a power-law distribution of \ping{the burst flux} $N\propto F^{-0.76\pm0.10}$, above $>2.6\times 10^{-9}\ {\rm erg\ cm^{-2}\ s^{-1}}$, steeper than \ping{that of} the X-ray burst storm in 2020 (index of $-0.5\pm0.1$, \citealt{Younes+2020ApJL}). 
The average waiting time for X-ray bursts was $59.95\ \rm s$, and the highest burst rate was $0.013 \rm s^{-1}$, suggesting that the 2022's X-ray reactivation is much weaker than \ping{that occurred in} 2020.

We used the persistent X-ray luminosity to estimate the SED of a fall-back disk model. The SED can satisfy the NIR counterpart detected with HST under reasonable assumptions of the disk, but future MIR observation is required to test the disk model and constrain the disk properties. 
We also compare the observations with the magnetosphere origin model, which is also possible to explain the NIR emission. \comm{
Future searches for counterparts at longer wavelengths, as did by \cite{Hare+2024ApJ_4U0142JWST} for 4U 0142+61 with JWST,} will help to find out which physical mechanism produces the IR emission from SGR~J1935+2154.

\section*{Acknowledgments}

This research is based on observations with the Gran Telescopio Canarias (GTC), installed at the Spanish Observatorio del Roque de los Muchachos of the Instituto de Astrofísica de Canarias, on the island of La Palma, and NICER (NASA). This research has made use of data and software provided by the High Energy Astrophysics Science Archive Research Center (HEASARC), which is a service of the Astrophysics Science Division at NASA/GSFC. The computation
was made by using the facilities at the High-Performance
Computing Center of Collaborative Innovation Center of
Advanced Microstructures (Nanjing University).

\comm{We are grateful to the anonymous referee for their careful reading and insightful comments.} Y.-X.S. and P.Z. acknowledge the support from NSFC grant No.\ 12273010. B.-B.Z. and Z.-H.Z. acknowledge the support of the Program for Innovative Talents, Entrepreneur in Jiangsu (JSSCTD202139). Z.-H.Z. acknowledges the support of the science research grants from the China Manned Space Project with NO. CMS-CSST-2021-A08. B.-B. Z. acknowledges the support by the National Key Research and Development Programs of China (2022YFF0711404, 2022SKA0130102), the National SKA Program of China (2022SKA0130100), the National Natural Science Foundation of China (grant Nos. 11833003, U2038105, U1831135, 12121003, 12393811, 13001106), the science research grants from the China Manned Space Project with NO. CMS-CSST-2021-B11, the Fundamental Research Funds for the Central Universities. X.-D.L. acknowledges the National Key Research and Development Program of China (2021YFA0718500) and the
Natural Science Foundation of China under grant Nos. 12041301 and 12121003. P.W. acknowledges support from the National Natural Science Foundation of China (NSFC) Programs No.11988101, 12041303, the CAS Youth Interdisciplinary Team, the Youth Innovation Promotion Association CAS (id. 2021055), and the Cultivation Project for FAST Scientific Payoff and Research Achievement of CAMS-CAS. Y.-D.H. acknowledges financial support from INAF through the GRAWITA Large Program Grant (ID: 1.05.12.01.04).

\software HEASoft (v6.32.1), NICERDAS (v011a), Xspec \citep[v12.13.1][]{Arnaud1996}, IRAF \citep{IRAF1, IRAF2}, AstroImageJ \citep{astroImageJ}.

\begin{figure*}[htb]
    \centering
    \includegraphics[scale=1.0]{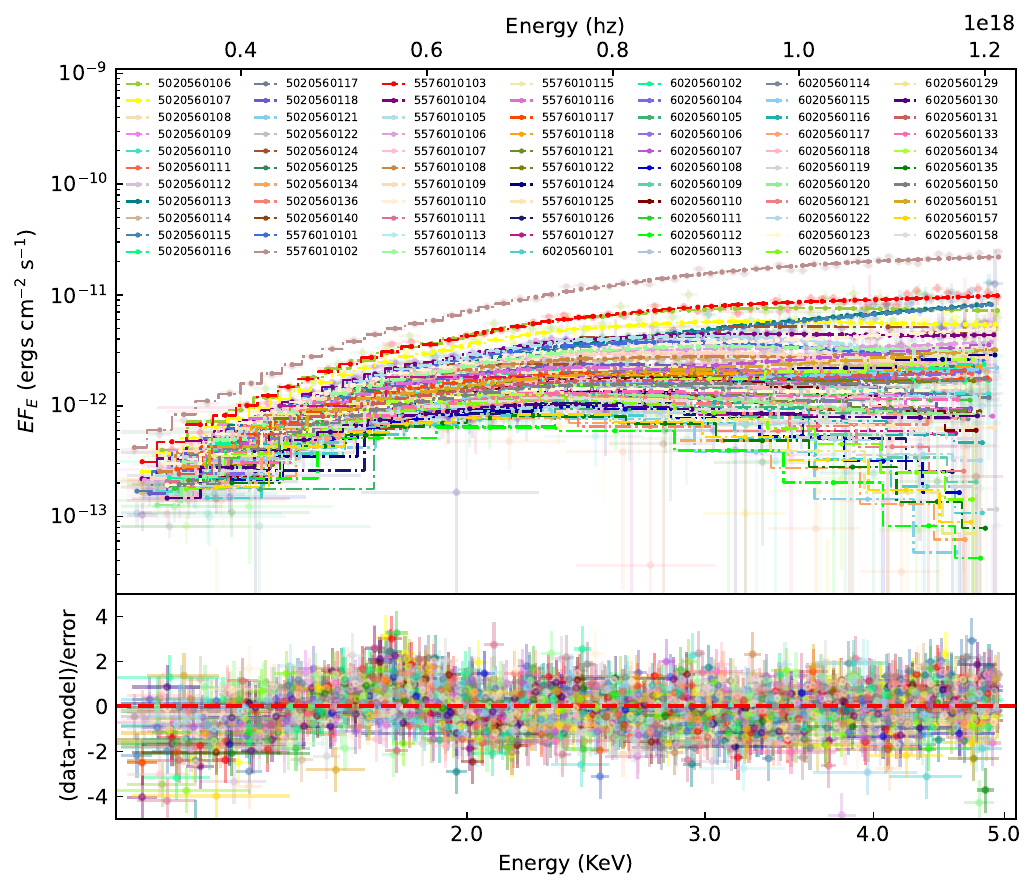}
    \caption{
    The persistent X-ray emission between 1.0--5.0 keV fitted using an absorbed (BB+PL) model. The spectra are displayed in $E F_{E}$ space.
    }
    \label{fig: persistent spectral}
\end{figure*}

\appendix
We fitted the 1--5 keV spectra for the persistent emission from SGR J1935+2154, with an absorbed BB+PL model as shown in Figure \ref{fig: persistent spectral}. The observation information and the fitting results are shown in Table \ref{tab: persistent spec}. We also fitted the X-ray bursts with an absorbed BB model, the information and the fitting results are presented in Table \ref{tab: burst spec}.

\begin{center}
    \setlength{\tabcolsep}{6mm}{
\begin{ThreePartTable}
\begin{longtable*}{ccccccc}
\caption{\textbf{Properties of the Persistent Emission}}\\
\label{tab: persistent spec}\\
\hline
\multirow{2}{*}{ObsID} & MJD & Exposure & Count Rate\tnote{a} & $kT$ & \multirow{2}{*}{Flux\tnote{b}} & BB Radius \\
&(days)&(s)&$\rm{(counts\ s^{-1})}$&(keV)&&(km)\\
\hline
\endfirsthead
\caption{ (Continued)}\\
\hline
\multirow{2}{*}{ObsID} & MJD & Exposure & Count Rate\tnote{a} & $kT$ & \multirow{2}{*}{Flux\tnote{b}} & BB Radius \\
&(days)&(s)&$\rm{(counts\ s^{-1})}$&(keV)&&(km)\\
\hline
\endhead
5020560106&59864.800&1192&$2.07\pm{0.04}$&${0.61}^{+0.04}_{-0.04}$&${2.02}^{+0.05}_{-0.05}$&${1.50}^{+0.14}_{-0.11}$\\ 

5020560107&59865.329&14527&$1.71\pm{0.01}$&${0.58}^{+0.01}_{-0.01}$&${1.61}^{+0.01}_{-0.01}$&${1.53}^{+0.04}_{-0.04}$\\ 

5576010101&59865.739&2698&$1.32\pm{0.02}$&${0.54}^{+0.02}_{-0.02}$&${1.00}^{+0.02}_{-0.02}$&${1.67}^{+0.11}_{-0.09}$\\ 

5576010102&59866.540&9592&$3.86\pm{0.02}$&${0.86}^{+0.02}_{-0.02}$&${4.67}^{+0.03}_{-0.03}$&${0.90}^{+0.05}_{-0.05}$\\ 

5576010103&59867.453&7695&$2.66\pm{0.02}$&${0.60}^{+0.02}_{-0.02}$&${2.47}^{+0.02}_{-0.02}$&${1.29}^{+0.07}_{-0.06}$\\ 

5576010104&59868.423&9914&$1.68\pm{0.01}$&${0.54}^{+0.02}_{-0.02}$&${1.33}^{+0.01}_{-0.01}$&${1.55}^{+0.07}_{-0.06}$\\ 

5576010105&59869.289&2590&$1.67\pm{0.03}$&${0.60}^{+0.03}_{-0.03}$&${0.93}^{+0.02}_{-0.02}$&${1.46}^{+0.09}_{-0.07}$\\ 

5576010106&59870.519&2434&$1.26\pm{0.02}$&${0.46}^{+0.03}_{-0.03}$&${1.16}^{+0.03}_{-0.03}$&${1.83}^{+0.23}_{-0.17}$\\
\hline

 \end{longtable*}
\begin{flushleft}
    \footnotesize
    \rep{\item[a] The count rate was computed after extracting the bursts, in the 1.0--5.0 keV interval.}
    \item[b] The unabsorbed flux between 0.3--10 keV in units of $(\rm{\times10^{-11}\ erg\ s^{-1}\ cm^{-2}\ })$
\end{flushleft}

\end{ThreePartTable}
}
\end{center}
\begin{center}
    \setlength{\tabcolsep}{1.4mm}{
\begin{longtable*}{cccccccccc}
\caption{\textbf{Burst Properties}}\\
\label{tab: burst spec}\\
\hline
\multirow{2}{*}{Burst ID} & $T_{\rm{start}}$ & $T_{\rm{peak}}$ & $T_{\rm{end}}$ & $T_{\rm{90}}$ &Count Rate\tnote{a} & $kT$ & Radius & ${\rm log_{10}} F$\tnote{b} &\multirow{2}{*}{$\chi^{2}/\rm d.o.f.$} \\
&(UTC)&(UTC)&(UTC)&(s)&$\rm{(counts\ s^{-1})}$&(keV)&(km)&($\rm{erg\ cm^{-2} \ s^{-1}}$)&\\
\hline
\endfirsthead
\caption{ (Continued)}\\
\hline
\multirow{2}{*}{Burst ID} & $T_{\rm{start}}$ & $T_{\rm{peak}}$ & $T_{\rm{end}}$ & $T_{\rm{90}}$ &Count Rate\tnote{a} & $kT$ & Radius & ${\rm log_{10}} F$\tnote{b} &\multirow{2}{*}{$\chi^{2}/\rm d.o.f.$} \\
&(UTC)&(UTC)&(UTC)&(s)&$\rm{(counts\ s^{-1})}$&(keV)&(km)&($\rm{erg\ cm^{-2}\ s^{-1}}$)&\\
\endhead

\hline 
 \multicolumn{9}{c}{ObsID: 5020560106 Date: 2022-10-12}\\ 
 \hline
1 &17:40:16.926&17:40:17.030&17:40:17.886&0.78&$243.90\pm{16.99}$&$1.68_{-0.16}^{+0.21}$&$4.47_{-0.55}^{+0.68}$&$-8.47_{-0.03}^{+0.03}$&53.89/47\\ 
2 &17:40:27.157&17:40:27.405&17:40:28.405&0.87&$1254.29\pm{33.86}$&$2.54_{-0.16}^{+0.18}$&$6.45_{-0.41}^{+0.45}$&$-7.60_{-0.01}^{+0.01}$&61.67/60\\ 
3 &17:41:18.587&17:41:18.651&17:41:18.819&0.15&$298.21\pm{37.28}$&$2.85_{-0.73}^{+1.69}$&$2.80_{-0.76}^{+1.16}$&$-8.19_{-0.06}^{+0.05}$&54.91/44\\ 
4 &17:44:12.234&17:44:12.330&17:44:12.666&0.13&$5757.13\pm{121.86}$&$5.78_{-0.69}^{+0.94}$&$7.03_{-0.66}^{+0.72}$&$-6.75_{-0.01}^{+0.01}$&86.68/63\\ 
5 &17:45:06.721&17:45:06.753&17:45:06.977&0.15&$199.67\pm{29.13}$&$4.26_{-1.66}^{+11.67}$&$1.64_{-0.67}^{+1.13}$&$-8.26_{-0.07}^{+0.06}$&44.24/43\\ 
6 &17:46:44.831&17:46:44.927&17:46:45.279&0.35&$84.67\pm{14.52}$&$1.28_{-0.22}^{+0.33}$&$3.77_{-0.88}^{+1.46}$&$-9.05_{-0.08}^{+0.07}$&50.60/43\\ 
7 &19:18:53.169&19:18:53.281&19:18:53.329&0.16&$91.92\pm{24.57}$&$1.77_{-0.54}^{+1.56}$&$2.58_{-0.94}^{+2.15}$&$-8.87_{-0.12}^{+0.11}$&41.10/43\\
\hline
 \end{longtable*}
 \begin{flushleft}
    \footnotesize
    \rep{\item[a] The count rate was computed in the 1.0--7.0 keV interval.}
    
    \item[b] The unabsorbed flux between 0.3--10 keV in units of $(\rm{\times10^{-11}\ erg\ s^{-1}\ cm^{-2}\ })$
\end{flushleft}
}
\end{center}

\bibliography{sample631}{}

\begin{thebibliography}{}
\expandafter\ifx\csname natexlab\endcsname\relax\def\natexlab#1{#1}\fi
\providecommand{\url}[1]{\href{#1}{#1}}
\providecommand{\dodoi}[1]{doi:~\href{http://doi.org/#1}{\nolinkurl{#1}}}
\providecommand{\doeprint}[1]{\href{http://ascl.net/#1}{\nolinkurl{http://ascl.net/#1}}}
\providecommand{\doarXiv}[1]{\href{https://arxiv.org/abs/#1}{\nolinkurl{https://arxiv.org/abs/#1}}}

\bibitem[{{Alpar}(2001)}]{Alpar+2001ApJ}
{Alpar}, M.~A. 2001, \apj, 554, 1245, \dodoi{10.1086/321393}

\bibitem[{{Arnaud}(1996)}]{Arnaud1996}
{Arnaud}, K.~A. 1996, in Astronomical Society of the Pacific Conference Series,
  Vol. 101, Astronomical Data Analysis Software and Systems V, ed. G.~H.
  {Jacoby} \& J.~{Barnes}, 17

\bibitem[{{Aschwanden}(2013)}]{Aschwanden+2013}
{Aschwanden}, M.~J. 2013, {Self-Organized Criticality Systems}

\bibitem[{{Beloborodov} \& {Thompson}(2007)}]{Beloborodov+2007ApJ}
{Beloborodov}, A.~M., \& {Thompson}, C. 2007, \apj, 657, 967,
  \dodoi{10.1086/508917}

\bibitem[{{Bochenek} {et~al.}(2020){Bochenek}, {Ravi}, {Belov}, {Hallinan},
  {Kocz}, {Kulkarni}, \& {McKenna}}]{Bochenek+2020Natur}
{Bochenek}, C.~D., {Ravi}, V., {Belov}, K.~V., {et~al.} 2020, \nat, 587, 59,
  \dodoi{10.1038/s41586-020-2872-x}

\bibitem[{{Borghese} {et~al.}(2020){Borghese}, {Coti Zelati}, {Rea},
  {Esposito}, {Israel}, {Mereghetti}, \& {Tiengo}}]{Borghese+2020ApJL}
{Borghese}, A., {Coti Zelati}, F., {Rea}, N., {et~al.} 2020, \apjl, 902, L2,
  \dodoi{10.3847/2041-8213/aba82a}

\bibitem[{{Borghese} {et~al.}(2022){Borghese}, {Coti Zelati}, {Israel},
  {Pilia}, {Burgay}, {Trudu}, {Zane}, {Turolla}, {Rea}, {Esposito},
  {Mereghetti}, {Tiengo}, \& {Possenti}}]{Borghese+2022MNRAS}
{Borghese}, A., {Coti Zelati}, F., {Israel}, G.~L., {et~al.} 2022, \mnras, 516,
  602, \dodoi{10.1093/mnras/stac1314}

\bibitem[{{Cepa} {et~al.}(2000){Cepa}, {Aguiar}, {Escalera},
  {Gonzalez-Serrano}, {Joven-Alvarez}, {Peraza}, {Rasilla}, {Rodriguez-Ramos},
  {Gonzalez}, {Cobos Duenas}, {Sanchez}, {Tejada}, {Bland-Hawthorn},
  {Militello}, \& {Rosa}}]{Cepa+2000}
{Cepa}, J., {Aguiar}, M., {Escalera}, V.~G., {et~al.} 2000, in Society of
  Photo-Optical Instrumentation Engineers (SPIE) Conference Series, Vol. 4008,
  Optical and IR Telescope Instrumentation and Detectors, ed. M.~{Iye} \& A.~F.
  {Moorwood}, 623--631, \dodoi{10.1117/12.395520}

\bibitem[{{Chatterjee} {et~al.}(2000){Chatterjee}, {Hernquist}, \&
  {Narayan}}]{Chatterjee+2000}
{Chatterjee}, P., {Hernquist}, L., \& {Narayan}, R. 2000, \apj, 534, 373,
  \dodoi{10.1086/308748}

\bibitem[{{Cheng} {et~al.}(1996){Cheng}, {Epstein}, {Guyer}, \&
  {Young}}]{Cheng+1996Natur}
{Cheng}, B., {Epstein}, R.~I., {Guyer}, R.~A., \& {Young}, A.~C. 1996, \nat,
  382, 518, \dodoi{10.1038/382518a0}

\bibitem[{{CHIME/FRB Collaboration} {et~al.}(2020){CHIME/FRB Collaboration},
  {Andersen}, {Bandura}, {Bhardwaj}, {Bij}, {Boyce}, {Boyle}, {Brar},
  {Cassanelli}, {Chawla}, {Chen}, {Cliche}, {Cook}, {Cubranic}, {Curtin},
  {Denman}, {Dobbs}, {Dong}, {Fandino}, {Fonseca}, {Gaensler}, {Giri}, {Good},
  {Halpern}, {Hill}, {Hinshaw}, {H{\"o}fer}, {Josephy}, {Kania}, {Kaspi},
  {Landecker}, {Leung}, {Li}, {Lin}, {Masui}, {McKinven}, {Mena-Parra},
  {Merryfield}, {Meyers}, {Michilli}, {Milutinovic}, {Mirhosseini},
  {M{\"u}nchmeyer}, {Naidu}, {Newburgh}, {Ng}, {Patel}, {Pen},
  {Pinsonneault-Marotte}, {Pleunis}, {Quine}, {Rafiei-Ravandi}, {Rahman},
  {Ransom}, {Renard}, {Sanghavi}, {Scholz}, {Shaw}, {Shin}, {Siegel}, {Singh},
  {Smegal}, {Smith}, {Stairs}, {Tan}, {Tendulkar}, {Tretyakov}, {Vanderlinde},
  {Wang}, {Wulf}, \& {Zwaniga}}]{CHIME+2020Natur}
{CHIME/FRB Collaboration}, {Andersen}, B.~C., {Bandura}, K.~M., {et~al.} 2020,
  \nat, 587, 54, \dodoi{10.1038/s41586-020-2863-y}

\bibitem[{{Chrimes} {et~al.}(2022{\natexlab{a}}){Chrimes}, {Levan}, {Fruchter},
  {Groot}, {Kouveliotou}, {Lyman}, {Tanvir}, \&
  {Wiersema}}]{Chrimes+2022HST_counterpart}
{Chrimes}, A.~A., {Levan}, A.~J., {Fruchter}, A.~S., {et~al.}
  2022{\natexlab{a}}, \mnras, 512, 6093, \dodoi{10.1093/mnras/stac870}

\bibitem[{{Chrimes} {et~al.}(2022{\natexlab{b}}){Chrimes}, {Levan}, {Fruchter},
  {Groot}, {Jonker}, {Kouveliotou}, {Lyman}, {Stanway}, {Tanvir}, \&
  {Wiersema}}]{Chrimes+2022MNRAS}
---. 2022{\natexlab{b}}, \mnras, 513, 3550, \dodoi{10.1093/mnras/stac1090}

\bibitem[{{Collins} {et~al.}(2017){Collins}, {Kielkopf}, {Stassun}, \&
  {Hessman}}]{astroImageJ}
{Collins}, K.~A., {Kielkopf}, J.~F., {Stassun}, K.~G., \& {Hessman}, F.~V.
  2017, \aj, 153, 77, \dodoi{10.3847/1538-3881/153/2/77}

\bibitem[{{Coti Zelati} {et~al.}(2018){Coti Zelati}, {Rea}, {Pons}, {Campana},
  \& {Esposito}}]{Coti+2018MNRAS}
{Coti Zelati}, F., {Rea}, N., {Pons}, J.~A., {Campana}, S., \& {Esposito}, P.
  2018, \mnras, 474, 961, \dodoi{10.1093/mnras/stx2679}

\bibitem[{{Cutri} {et~al.}(2003){Cutri}, {Skrutskie}, {van Dyk}, {Beichman},
  {Carpenter}, {Chester}, {Cambresy}, {Evans}, {Fowler}, {Gizis}, {Howard},
  {Huchra}, {Jarrett}, {Kopan}, {Kirkpatrick}, {Light}, {Marsh}, {McCallon},
  {Schneider}, {Stiening}, {Sykes}, {Weinberg}, {Wheaton}, {Wheelock}, \&
  {Zacarias}}]{2mass+2003}
{Cutri}, R.~M., {Skrutskie}, M.~F., {van Dyk}, S., {et~al.} 2003, {2MASS All
  Sky Catalog of point sources.}

\bibitem[{{Dhillon} {et~al.}(2011){Dhillon}, {Marsh}, {Littlefair},
  {Copperwheat}, {Hickman}, {Kerry}, {Levan}, {Rea}, {Savoury}, {Tanvir},
  {Turolla}, \& {Wiersema}}]{Dhillon+2011}
{Dhillon}, V.~S., {Marsh}, T.~R., {Littlefair}, S.~P., {et~al.} 2011, \mnras,
  416, L16, \dodoi{10.1111/j.1745-3933.2011.01088.x}

\bibitem[{{Dong} \& {Chime/Frb Collaboration}(2022)}]{2022ATelDong}
{Dong}, F.~A., \& {Chime/Frb Collaboration}. 2022, The Astronomer's Telegram,
  15681, 1

\bibitem[{{Duncan} \& {Thompson}(1992)}]{Duncan+1992ApJ}
{Duncan}, R.~C., \& {Thompson}, C. 1992, \apjl, 392, L9, \dodoi{10.1086/186413}

\bibitem[{{Esposito} {et~al.}(2021){Esposito}, {Rea}, \&
  {Israel}}]{Esposito+2021ASSL}
{Esposito}, P., {Rea}, N., \& {Israel}, G.~L. 2021, in Astrophysics and Space
  Science Library, Vol. 461, Timing Neutron Stars: Pulsations, Oscillations and
  Explosions, ed. T.~M. {Belloni}, M.~{M{\'e}ndez}, \& C.~{Zhang}, 97--142,
  \dodoi{10.1007/978-3-662-62110-3_3}

\bibitem[{{Foight} {et~al.}(2016){Foight}, {G{\"u}ver}, {{\"O}zel}, \&
  {Slane}}]{Foight+2016ApJ}
{Foight}, D.~R., {G{\"u}ver}, T., {{\"O}zel}, F., \& {Slane}, P.~O. 2016, \apj,
  826, 66, \dodoi{10.3847/0004-637X/826/1/66}

\bibitem[{{Frank} {et~al.}(2002){Frank}, {King}, \&
  {Raine}}]{Frank+2002apa..book.....F}
{Frank}, J., {King}, A., \& {Raine}, D.~J. 2002, {Accretion Power in
  Astrophysics: Third Edition}

\bibitem[{{Gavriil} {et~al.}(2004){Gavriil}, {Kaspi}, \&
  {Woods}}]{Gavriil+2004ApJ...607..959G}
{Gavriil}, F.~P., {Kaspi}, V.~M., \& {Woods}, P.~M. 2004, \apj, 607, 959,
  \dodoi{10.1086/383564}

\bibitem[{{Gendreau} {et~al.}(2016){Gendreau}, {Arzoumanian}, {Adkins},
  {Albert}, {Anders}, {Aylward}, {Baker}, {Balsamo}, {Bamford}, {Benegalrao},
  {Berry}, {Bhalwani}, {Black}, {Blaurock}, {Bronke}, {Brown}, {Budinoff},
  {Cantwell}, {Cazeau}, {Chen}, {Clement}, {Colangelo}, {Coleman},
  {Coopersmith}, {Dehaven}, {Doty}, {Egan}, {Enoto}, {Fan}, {Ferro}, {Foster},
  {Galassi}, {Gallo}, {Green}, {Grosh}, {Ha}, {Hasouneh}, {Heefner}, {Hestnes},
  {Hoge}, {Jacobs}, {J{\o}rgensen}, {Kaiser}, {Kellogg}, {Kenyon}, {Koenecke},
  {Kozon}, {LaMarr}, {Lambertson}, {Larson}, {Lentine}, {Lewis}, {Lilly},
  {Liu}, {Malonis}, {Manthripragada}, {Markwardt}, {Matonak}, {Mcginnis},
  {Miller}, {Mitchell}, {Mitchell}, {Mohammed}, {Monroe}, {Montt de Garcia},
  {Mul{\'e}}, {Nagao}, {Ngo}, {Norris}, {Norwood}, {Novotka}, {Okajima},
  {Olsen}, {Onyeachu}, {Orosco}, {Peterson}, {Pevear}, {Pham}, {Pollard},
  {Pope}, {Powers}, {Powers}, {Price}, {Prigozhin}, {Ramirez}, {Reid},
  {Remillard}, {Rogstad}, {Rosecrans}, {Rowe}, {Sager}, {Sanders}, {Savadkin},
  {Saylor}, {Schaeffer}, {Schweiss}, {Semper}, {Serlemitsos}, {Shackelford},
  {Soong}, {Struebel}, {Vezie}, {Villasenor}, {Winternitz}, {Wofford},
  {Wright}, {Yang}, \& {Yu}}]{Gendreau+2016}
{Gendreau}, K.~C., {Arzoumanian}, Z., {Adkins}, P.~W., {et~al.} 2016, in
  Society of Photo-Optical Instrumentation Engineers (SPIE) Conference Series,
  Vol. 9905, Space Telescopes and Instrumentation 2016: Ultraviolet to Gamma
  Ray, ed. J.-W.~A. {den Herder}, T.~{Takahashi}, \& M.~{Bautz}, 99051H,
  \dodoi{10.1117/12.2231304}

\bibitem[{{Hare} {et~al.}(2024){Hare}, {Pavlov}, {Posselt}, {Kargaltsev},
  {Temim}, \& {Chen}}]{Hare+2024ApJ_4U0142JWST}
{Hare}, J., {Pavlov}, G.~G., {Posselt}, B., {et~al.} 2024, \apj, 972, 176,
  \dodoi{10.3847/1538-4357/ad5ce5}

\bibitem[{{Hulleman} {et~al.}(2000){Hulleman}, {van Kerkwijk}, \&
  {Kulkarni}}]{Hulleman+2000Natur}
{Hulleman}, F., {van Kerkwijk}, M.~H., \& {Kulkarni}, S.~R. 2000, \nat, 408,
  689, \dodoi{10.1038/35047024}

\bibitem[{{Israel} {et~al.}(2016){Israel}, {Esposito}, {Rea}, {Coti Zelati},
  {Tiengo}, {Campana}, {Mereghetti}, {Rodriguez Castillo}, {G{\"o}tz},
  {Burgay}, {Possenti}, {Zane}, {Turolla}, {Perna}, {Cannizzaro}, \&
  {Pons}}]{Israel+2016MNRAS}
{Israel}, G.~L., {Esposito}, P., {Rea}, N., {et~al.} 2016, \mnras, 457, 3448,
  \dodoi{10.1093/mnras/stw008}

\bibitem[{{Kaspi} \& {Beloborodov}(2017)}]{Kaspi+2017ARA&A}
{Kaspi}, V.~M., \& {Beloborodov}, A.~M. 2017, \araa, 55, 261,
  \dodoi{10.1146/annurev-astro-081915-023329}

\bibitem[{{Levan} {et~al.}(2018){Levan}, {Kouveliotou}, \&
  {Fruchter}}]{Levan+2018ApJ}
{Levan}, A., {Kouveliotou}, C., \& {Fruchter}, A. 2018, \apj, 854, 161,
  \dodoi{10.3847/1538-4357/aaa88d}

\bibitem[{{Li} {et~al.}(2021){Li}, {Lin}, {Xiong}, {Ge}, {Li}, {Li}, {Lu},
  {Zhang}, {Tuo}, {Nang}, {Zhang}, {Xiao}, {Chen}, {Song}, {Xu}, {Liu}, {Jia},
  {Cao}, {Qu}, {Zhang}, {Gu}, {Liao}, {Zhao}, {Tan}, {Nie}, {Zhao}, {Zheng},
  {Zheng}, {Luo}, {Cai}, {Li}, {Xue}, {Bu}, {Chang}, {Chen}, {Chen}, {Chen},
  {Chen}, {Chen}, {Cui}, {Cui}, {Deng}, {Dong}, {Du}, {Fu}, {Gao}, {Gao},
  {Gao}, {Gu}, {Guan}, {Guo}, {Han}, {Huang}, {Huo}, {Jiang}, {Jiang}, {Jin},
  {Jin}, {Kong}, {Li}, {Li}, {Li}, {Li}, {Li}, {Li}, {Li}, {Liang}, {Liu},
  {Liu}, {Liu}, {Liu}, {Liu}, {Lu}, {Lu}, {Luo}, {Ma}, {Meng}, {Ou}, {Sai},
  {Shang}, {Song}, {Sun}, {Tao}, {Wang}, {Wang}, {Wang}, {Wang}, {Wang}, {Wen},
  {Wu}, {Wu}, {Wu}, {Xiao}, {Xu}, {Yang}, {Yang}, {Yang}, {Yang}, {Yi}, {Yin},
  {You}, {Zhang}, {Zhang}, {Zhang}, {Zhang}, {Zhang}, {Zhang}, {Zhang},
  {Zhang}, {Zhang}, {Zhang}, {Zhang}, {Zhang}, {Zhang}, {Zhang}, {Zhang},
  {Zhang}, {Zhou}, {Zhou}, {Zhu}, {Zhu}, \& {Zhuang}}]{Li+2021NatAs}
{Li}, C.~K., {Lin}, L., {Xiong}, S.~L., {et~al.} 2021, Nature Astronomy, 5,
  378, \dodoi{10.1038/s41550-021-01302-6}

\bibitem[{{Lin} {et~al.}(2020){Lin}, {G{\"o}{\u{g}}{\"u}{\c{s}}}, {Roberts},
  {Kouveliotou}, {Kaneko}, {van der Horst}, \& {Younes}}]{lin+2020ApJ}
{Lin}, L., {G{\"o}{\u{g}}{\"u}{\c{s}}}, E., {Roberts}, O.~J., {et~al.} 2020,
  \apj, 893, 156, \dodoi{10.3847/1538-4357/ab818f}

\bibitem[{{Lin} {et~al.}(2013){Lin}, {G{\"o}{\v{g}}{\"u}{\c{s}}}, {Kaneko}, \&
  {Kouveliotou}}]{Lin+2013ApJ}
{Lin}, L., {G{\"o}{\v{g}}{\"u}{\c{s}}}, E., {Kaneko}, Y., \& {Kouveliotou}, C.
  2013, \apj, 778, 105, \dodoi{10.1088/0004-637X/778/2/105}

\bibitem[{{Lyman} {et~al.}(2022){Lyman}, {Levan}, {Wiersema}, {Kouveliotou},
  {Chrimes}, \& {Fruchter}}]{Lyman+2022ApJ}
{Lyman}, J.~D., {Levan}, A.~J., {Wiersema}, K., {et~al.} 2022, \apj, 926, 121,
  \dodoi{10.3847/1538-4357/ac432f}

\bibitem[{{Maan} {et~al.}(2022){Maan}, {Straal}, {van Leeuwen}, \&
  {Pastor-Marazuela}}]{2022AtelMaan}
{Maan}, Y., {Straal}, S., {van Leeuwen}, J., \& {Pastor-Marazuela}, I. 2022,
  The Astronomer's Telegram, 15168, 1

\bibitem[{{Olausen} \& {Kaspi}(2014)}]{Olausen+2014ApJS}
{Olausen}, S.~A., \& {Kaspi}, V.~M. 2014, \apjs, 212, 6,
  \dodoi{10.1088/0067-0049/212/1/6}

\bibitem[{{Palm}(2022{\natexlab{a}})}]{Palm+2022ATel}
{Palm}, D.~M. 2022{\natexlab{a}}, The Astronomer's Telegram, 15667, 1

\bibitem[{{Palm}(2022{\natexlab{b}})}]{2022ATelPalm}
---. 2022{\natexlab{b}}, The Astronomer's Telegram, 15667, 1

\bibitem[{{Pearlman} \& {Chime/Frb Collaboration}(2022)}]{2022ATelCHIME}
{Pearlman}, A.~B., \& {Chime/Frb Collaboration}. 2022, The Astronomer's
  Telegram, 15792, 1

\bibitem[{{Perna} {et~al.}(2000){Perna}, {Hernquist}, \&
  {Narayan}}]{Perna+2000ApJ}
{Perna}, R., {Hernquist}, L., \& {Narayan}, R. 2000, \apj, 541, 344,
  \dodoi{10.1086/309404}

\bibitem[{{Remillard} {et~al.}(2022){Remillard}, {Loewenstein}, {Steiner},
  {Prigozhin}, {LaMarr}, {Enoto}, {Gendreau}, {Arzoumanian}, {Markwardt},
  {Basak}, {Stevens}, {Ray}, {Altamirano}, \& {Buisson}}]{Remillard+2022AJ3C50}
{Remillard}, R.~A., {Loewenstein}, M., {Steiner}, J.~F., {et~al.} 2022, \aj,
  163, 130, \dodoi{10.3847/1538-3881/ac4ae6}

\bibitem[{{Roberts}(2022)}]{2022ATelRoberts}
{Roberts}, O.~J. 2022, The Astronomer's Telegram, 15162, 1

\bibitem[{{Smith} {et~al.}(2002){Smith}, {Tucker}, {Kent}, {Richmond},
  {Fukugita}, {Ichikawa}, {Ichikawa}, {Jorgensen}, {Uomoto}, {Gunn}, {Hamabe},
  {Watanabe}, {Tolea}, {Henden}, {Annis}, {Pier}, {McKay}, {Brinkmann}, {Chen},
  {Holtzman}, {Shimasaku}, \& {York}}]{Smith+2002AJ}
{Smith}, J.~A., {Tucker}, D.~L., {Kent}, S., {et~al.} 2002, \aj, 123, 2121,
  \dodoi{10.1086/339311}

\bibitem[{{Stamatikos} {et~al.}(2014){Stamatikos}, {Malesani}, {Page}, \&
  {Sakamoto}}]{Stamatikos+2014GCN}
{Stamatikos}, M., {Malesani}, D., {Page}, K.~L., \& {Sakamoto}, T. 2014, GRB
  Coordinates Network, 16520, 1

\bibitem[{{Tody}(1986)}]{IRAF1}
{Tody}, D. 1986, in Society of Photo-Optical Instrumentation Engineers (SPIE)
  Conference Series, Vol. 627, Instrumentation in astronomy VI, ed. D.~L.
  {Crawford}, 733, \dodoi{10.1117/12.968154}

\bibitem[{{Tody}(1993)}]{IRAF2}
{Tody}, D. 1993, in Astronomical Society of the Pacific Conference Series,
  Vol.~52, Astronomical Data Analysis Software and Systems II, ed. R.~J.
  {Hanisch}, R.~J.~V. {Brissenden}, \& J.~{Barnes}, 173

\bibitem[{{Turolla} {et~al.}(2015){Turolla}, {Zane}, \&
  {Watts}}]{Turolla+2015RPPh}
{Turolla}, R., {Zane}, S., \& {Watts}, A.~L. 2015, Reports on Progress in
  Physics, 78, 116901, \dodoi{10.1088/0034-4885/78/11/116901}

\bibitem[{{Verner} {et~al.}(1996){Verner}, {Ferland}, {Korista}, \&
  {Yakovlev}}]{Verner+1996ApJ}
{Verner}, D.~A., {Ferland}, G.~J., {Korista}, K.~T., \& {Yakovlev}, D.~G. 1996,
  \apj, 465, 487, \dodoi{10.1086/177435}

\bibitem[{{Vrtilek} {et~al.}(1990){Vrtilek}, {Raymond}, {Garcia}, {Verbunt},
  {Hasinger}, \& {Kurster}}]{Vrtilek+1990A&A}
{Vrtilek}, S.~D., {Raymond}, J.~C., {Garcia}, M.~R., {et~al.} 1990, \aap, 235,
  162

\bibitem[{{Wang} {et~al.}(2023){Wang}, {Li}, {Ji}, {Hou}, {Gugercinoglu}, {Li},
  {Torres}, {Chen}, {Niu}, {Zhu}, {Zhang}, {Liang}, {Zhang}, {Ge}, {Dai},
  {Lin}, {Han}, {Feng}, {Niu}, {Zhang}, {Zhou}, {Xu}, {Zhang}, {Jiang}, {Miao},
  {Yuan}, {Wang}, {Yue}, {Wu}, {Wang}, {Wang}, {Gan}, {Li}, {Sun}, \&
  {Chi}}]{Wangpei+23}
{Wang}, P., {Li}, J., {Ji}, L., {et~al.} 2023, arXiv e-prints,
  arXiv:2308.08832, \dodoi{10.48550/arXiv.2308.08832}

\bibitem[{{Wang} \& {Chen}(2019)}]{Wang+2019ApJ}
{Wang}, S., \& {Chen}, X. 2019, \apj, 877, 116,
  \dodoi{10.3847/1538-4357/ab1c61}

\bibitem[{{Wang} {et~al.}(2008){Wang}, {Bassa}, {Kaspi}, {Bryant}, \&
  {Morrell}}]{Wang+2008ApJ}
{Wang}, Z., {Bassa}, C., {Kaspi}, V.~M., {Bryant}, J.~J., \& {Morrell}, N.
  2008, \apj, 679, 1443, \dodoi{10.1086/587505}

\bibitem[{{Wang} {et~al.}(2006){Wang}, {Chakrabarty}, \&
  {Kaplan}}]{Wang+2006Natur}
{Wang}, Z., {Chakrabarty}, D., \& {Kaplan}, D.~L. 2006, \nat, 440, 772,
  \dodoi{10.1038/nature04669}

\bibitem[{{Wilms} {et~al.}(2000){Wilms}, {Allen}, \& {McCray}}]{Wilms+2000}
{Wilms}, J., {Allen}, A., \& {McCray}, R. 2000, \apj, 542, 914,
  \dodoi{10.1086/317016}

\bibitem[{{Younes} {et~al.}(2022){Younes}, {Burns}, {Roberts}, {Wood}, {Veres},
  \& {Kouveliotou}}]{Younes+2022ATel}
{Younes}, G., {Burns}, E., {Roberts}, O.~J., {et~al.} 2022, The Astronomer's
  Telegram, 15794, 1

\bibitem[{{Younes} {et~al.}(2017){Younes}, {Kouveliotou}, {Jaodand}, {Baring},
  {van der Horst}, {Harding}, {Hessels}, {Gehrels}, {Gill}, {Huppenkothen},
  {Granot}, {G{\"o}{\u{g}}{\"u}{\c{s}}}, \& {Lin}}]{Younes+2017ApJ}
{Younes}, G., {Kouveliotou}, C., {Jaodand}, A., {et~al.} 2017, \apj, 847, 85,
  \dodoi{10.3847/1538-4357/aa899a}

\bibitem[{{Younes} {et~al.}(2020){Younes}, {G{\"u}ver}, {Kouveliotou},
  {Baring}, {Hu}, {Wadiasingh}, {Begi{\c{c}}arslan}, {Enoto},
  {G{\"o}{\u{g}}{\"u}{\c{s}}}, {Lin}, {Harding}, {van der Horst}, {Majid},
  {Guillot}, \& {Malacaria}}]{Younes+2020ApJL}
{Younes}, G., {G{\"u}ver}, T., {Kouveliotou}, C., {et~al.} 2020, \apjl, 904,
  L21, \dodoi{10.3847/2041-8213/abc94c}

\bibitem[{{Younes} {et~al.}(2023){Younes}, {Baring}, {Harding}, {Enoto},
  {Wadiasingh}, {Pearlman}, {Ho}, {Guillot}, {Arzoumanian}, {Borghese},
  {Gendreau}, {G{\"o}{\v{g}}{\"u}{\c{s}}}, {G{\"u}ver}, {van der Horst}, {Hu},
  {Jaisawal}, {Kouveliotou}, {Lin}, \& {Majid}}]{Younes+2023NatAs}
{Younes}, G., {Baring}, M.~G., {Harding}, A.~K., {et~al.} 2023, Nature
  Astronomy, 7, 339, \dodoi{10.1038/s41550-022-01865-y}

\bibitem[{{Zane} {et~al.}(2011){Zane}, {Nobili}, \& {Turolla}}]{Zane+2011ASSP}
{Zane}, S., {Nobili}, L., \& {Turolla}, R. 2011, in Astrophysics and Space
  Science Proceedings, Vol.~21, High-Energy Emission from Pulsars and their
  Systems, 329, \dodoi{10.1007/978-3-642-17251-9_26}

\bibitem[{{Zhou} {et~al.}(2020){Zhou}, {Zhou}, {Chen}, {Wang}, {Vink}, \&
  {Wang}}]{Zhou+2020ApJ}
{Zhou}, P., {Zhou}, X., {Chen}, Y., {et~al.} 2020, \apj, 905, 99,
  \dodoi{10.3847/1538-4357/abc34a}

\bibitem[{{Zhu} {et~al.}(2023){Zhu}, {Xu}, {Zhou}, {Lin}, {Wang}, {Wang},
  {Zhang}, {Niu}, {Chen}, {Li}, {Meng}, {Lee}, {Zhang}, {Feng}, {Ge},
  {G{\"o}{\u{g}}{\"u}{\c{s}}}, {Guan}, {Han}, {Jiang}, {Jiang}, {Kouveliotou},
  {Li}, {Miao}, {Miao}, {Men}, {Niu}, {Wang}, {Wang}, {Xu}, {Xu}, {Xue},
  {Yang}, {Yu}, {Yuan}, {Yue}, {Zhang}, \& {Zhang}}]{ZhuWeiWei+2023SciA}
{Zhu}, W., {Xu}, H., {Zhou}, D., {et~al.} 2023, Science Advances, 9, eadf6198,
  \dodoi{10.1126/sciadv.adf6198}

\end{thebibliography}
\bibliographystyle{aasjournal}

\end{CJK*}
\end{document}